\documentclass[pdftex,colorlinks,notitlepage,a4paper,12pt]{article}

\usepackage{slantsc}

\usepackage[a4paper,margin=1in]{geometry}

\usepackage[pdftex]{graphicx}

\usepackage{epstopdf}

\usepackage[usenames,dvipsnames]{color}
\usepackage[dvipsnames]{xcolor}

\usepackage{soul}
\soulregister\parencite7
\soulregister\textcite7
\soulregister\cite7
\soulregister\posscite7
\soulregister\ref7
\soulregister\enquote7

\newlength{\mylength}
\makeatletter
\newcommand{\mycfs}[1]{%
	\normalsize
	\@defaultunits\mylength=#1pt\relax\@nnil
	\edef\@tempa{{\strip@pt\mylength}}%
	\ifx\protect\@typeset@protect
	\edef\@currsize{\noexpand\mycfs\@tempa}
	\fi
	\mylength=1.2\mylength
	\edef\@tempa{\@tempa{\strip@pt\mylength}}%
	\expandafter\fontsize\@tempa
	\selectfont
}

\usepackage[utf8]{inputenc}
\usepackage[T1]{fontenc}

\PassOptionsToPackage{noconfigs}{babel}
\usepackage[french, spanish, british]{babel}
\babeltags{es = spanish}

\usepackage[style=american]{csquotes}

\usepackage[en-GB]{datetime2}
\DTMlangsetup[en-GB]{ord=raise}

\usepackage{textgreek}

\usepackage{latexsym}

\usepackage{amsmath}

\usepackage{econometrics}

\usepackage{amssymb}

\usepackage{bbold}

\usepackage{pdflscape} 

\usepackage{afterpage}

\usepackage{array}
\usepackage{delarray}
\usepackage{cases}
\usepackage{rotating}
\usepackage{longtable}
\usepackage{threeparttable}
\usepackage{threeparttablex}
\usepackage{booktabs}
\usepackage{caption}
\usepackage{subcaption}
\usepackage{makecell}

\usepackage{tabularx}
\usepackage{ltablex}
\keepXColumns
\usepackage{tabu}
\usepackage{multirow}

\makeatletter
\newlength\LongtableWidth
\newcommand*{\org@longtable}{}
\let\org@longtable\longtable
\def\longtable{%
	\begingroup
	\advance\c@LT@tables\@ne
	\edef\x{LT@\romannumeral\c@LT@tables}
	\global\LongtableWidth\z@
	\@ifundefined{\x}{%
	}{%
		\def\LT@entry##1##2{%
			\global\advance\LongtableWidth##2\relax
		}%
		\@nameuse{\x}%
	}%
	\typeout{* \x: \the\LongtableWidth}%
	\endgroup \ifdim\LongtableWidth>\z@ \setlength{\LTcapwidth}{\LongtableWidth}%
	\fi
	\org@longtable }
\makeatother

\usepackage{ragged2e}

\usepackage{xpatch}

\usepackage{float}
\usepackage{placeins}

\usepackage{eurosym}    

\usepackage{appendix}

\usepackage[hyphens]{url}

\usepackage{xurl}

\hyphenchar\font=\string"7F
\usepackage[super]{nth}

\usepackage[shortlabels]{enumitem} 
\setlist[enumerate]{nosep, label = {(\arabic*)}}
\setlist[itemize]{nosep, label = {---}}

\usepackage{titlesec}
\titlelabel{\thetitle.\quad}
\titleformat*{\section}{\normalfont\large\bfseries\setstretch{1.5}}
\titleformat*{\subsection}{\normalfont\large\bfseries\setstretch{1}}

\usepackage{fancyhdr}

\makeatletter
\newread\pin@file
\newcounter{pinlineno}
\newcommand\pin@accu{}
\newcommand\pin@ext{pintmp}
\newcommand*\partialinput [3] {%
	\IfFileExists{#3}{%
		\openin\pin@file #3
		\setcounter{pinlineno}{1}
		\@whilenum\value{pinlineno}<#1 \do{%
			\read\pin@file to\pin@line
			\stepcounter{pinlineno}%
		}
		\addtocounter{pinlineno}{-1}
		\let\pin@accu\empty
		\begingroup
		\endlinechar\newlinechar
		\@whilenum\value{pinlineno}<#2 \do{%
			\readline\pin@file to\pin@line
			\edef\pin@accu{\pin@accu\pin@line}%
			\stepcounter{pinlineno}%
		}
		\closein\pin@file
		\expandafter\endgroup
		\scantokens\expandafter{\pin@accu}%
	}{%
		\errmessage{File `#3' doesn't exist!}%
	}%
}
\makeatother

\usepackage[backend=biber,style=apa,uniquelist=false,uniquename=false,dashed=true,dateabbrev=false,eventdate=comp,hyperref=true,language=british,parentracker=true]{biblatex}

\DeclareNameFormat{labelname:poss}{
	\nameparts{#1}
	\ifcase\value{uniquename}%
	\usebibmacro{name:family}{\namepartfamily}{\namepartgiven}{\namepartprefix}{\namepartsuffix}%
	\or
	\ifuseprefix
	{\usebibmacro{name:first-last}{\namepartfamily}{\namepartgiveni}{\namepartprefix}{\namepartsuffixi}}
	{\usebibmacro{name:first-last}{\namepartfamily}{\namepartgiveni}{\namepartprefixi}{\namepartsuffixi}}%
	\or
	\usebibmacro{name:first-last}{\namepartfamily}{\namepartgiven}{\namepartprefix}{\namepartsuffix}%
	\fi
	\usebibmacro{name:andothers}%
	\ifnumequal{\value{listcount}}{\value{liststop}}{'s}{}}
\DeclareFieldFormat{shorthand:poss}{%
	\ifnameundef{labelname}{#1's}{#1}}
\DeclareFieldFormat{citetitle:poss}{\mkbibemph{#1}'s}
\DeclareFieldFormat{label:poss}{#1's}
\newrobustcmd*{\posscitealias}{%
	\AtNextCite{%
		\DeclareNameAlias{labelname}{labelname:poss}%
		\DeclareFieldAlias{shorthand}{shorthand:poss}%
		\DeclareFieldAlias{citetitle}{citetitle:poss}%
		\DeclareFieldAlias{label}{label:poss}}}
\newrobustcmd*{\posscite}{%
	\posscitealias%
	\textcite}
\newrobustcmd*{\Posscite}{\bibsentence\posscite}
\newrobustcmd*{\posscites}{%
	\posscitealias%
	\textcites}

\DeclareFieldFormat*{edition}{\nth{#1} ed.}

\newcommand{\noop}[1]{}

\usepackage{microtype}

\usepackage{siunitx} 

\usepackage{setspace}
\usepackage[bottom]{footmisc}

\usepackage[framemethod=tikz]{mdframed}
\usepackage{tikz}
\usetikzlibrary{shapes.geometric, arrows, positioning, calc, decorations.markings, shapes.misc, fit}

\usepackage{acronym}

\usepackage{etoolbox}
\makeatletter
\patchcmd{\maketitle}{\@makefntext}{\fakecommand}{}{}
\patchcmd{\maketitle}{\rlap}{\hbox}{}{}
\patchcmd{\@maketitle}{\@author}{\hspace*{5pt}\@author}{}{}
\makeatother

\usepackage{comment}

\usepackage{changes}

\definechangesauthor[color=Green]{NA}

\usepackage{todonotes}

\usepackage{abstract}
\usepackage{titling}

\usepackage[noblocks]{authblk}

\title{The long-term impact of (un)conditional cash transfers on labour market outcomes in Ecuador\thanks{Corresponding author: Jos{\'e}-Ignacio Ant{\'o}n, Department of Applied Economics, University of Salamanca, Campus Miguel de Unamuno, 37007 Salamanca (Spain), e-mail: \href{mailto:janton@usal.es}{\texttt{janton@usal.es}}. We thank Zuleika Ferre, Rafael Mu{\~n}oz de Bustillo, Guillermo Orfao, Miguel Paradela-L{\'o}pez, Alberto del Rey, Sudipa Sarkar, Nicole Schneeweis, Patricia Triunfo and Rudolf Winter-Ebmer for helpful comments on a previous version of the paper. Ant{\'o}n acknowledges the financial support from the Spanish Ministry of Science, Innovation and Universities (research project PID2021-123875NB-I00).}}

\pretitle{\vspace{-40pt}\centering\Large\bfseries}
	\posttitle{\par\centering}

\predate{\centering}
	\postdate{\centering\vskip -2em}

\makeatletter
\AtBeginDocument{%
	\newlength{\temp@x}%
	\newlength{\temp@y}%
	\newlength{\temp@w}%
	\newlength{\temp@h}%
	\def\my@coords#1#2#3#4{%
		\setlength{\temp@x}{#1}%
		\setlength{\temp@y}{#2}%
		\setlength{\temp@w}{#3}%
		\setlength{\temp@h}{#4}%
		\adjustlengths{}%
		\my@pdfliteral{\strip@pt\temp@x\space\strip@pt\temp@y\space\strip@pt\temp@w\space\strip@pt\temp@h\space re}}%
	\ifpdf
	\typeout{In PDF mode}%
	\def\my@pdfliteral#1{\pdfliteral page{#1}}
	\def\adjustlengths{}%
	\fi
	\ifxetex
	\def\my@pdfliteral #1{}
	\def\adjustlengths{\setlength{\temp@h}{-\temp@h}\addtolength{\temp@y}{1in}\addtolength{\temp@x}{-1in}}%
	\fi%
	\def\Hy@colorlink#1{%
		\begingroup
		\ifHy@ocgcolorlinks
		\def\Hy@ocgcolor{#1}%
		\my@pdfliteral{q}%
		\my@pdfliteral{7 Tr}
		\else
		\HyColor@UseColor#1%
		\fi
	}%
	\def\Hy@endcolorlink{%
		\ifHy@ocgcolorlinks%
		\my@pdfliteral{/OC/OCPrint BDC}%
		\my@coords{0pt}{0pt}{\pdfpagewidth}{\pdfpageheight}%
		\my@pdfliteral{F}
		\my@pdfliteral{EMC/OC/OCView BDC}%
		\begingroup%
		\expandafter\HyColor@UseColor\Hy@ocgcolor%
		\my@coords{0pt}{0pt}{\pdfpagewidth}{\pdfpageheight}%
		\my@pdfliteral{F}
		\endgroup%
		\my@pdfliteral{EMC}%
		\my@pdfliteral{0 Tr}
		\my@pdfliteral{Q}%
		\fi
		\endgroup
	}%
}
\makeatother

\DeclareLabeldate{%
	\field{date}
	\field{eventdate}
	\field{origdate}
	\field{urldate}
	\field{pubstate}
	\literal{nodate}
}

\renewbibmacro*{addendum+pubstate}{%
	\printfield{addendum}%
	\iffieldequalstr{labeldatesource}{pubstate}{}
	{\newunit\newblock\printfield{pubstate}}}

\author[$\dag$]{Juan~Ponce}
\affil[$\dag$]{FLACSO-Ecuador (Ecuador)}
\author[$\ddag$]{Jos{\'e}-Ignacio~Ant{\'o}n}
\affil[$\ddag$]{University of Salamanca (Spain)}
\author[$\S$]{Mercedes~Onofa}
\affil[$\S$]{Universidad Tecnol{\'o}gica Equinocial (Ecuador)}
\author[$\P$]{Roberto~Castillo}
\affil[$\P$]{Instituto Nacional de Estad{\'i}stica y Censos (Ecuador)}

\usepackage[pdftex,pdftitle={},pdfsubject={},%
pdfauthor={},breaklinks=true]{hyperref}
\hypersetup{
	colorlinks=true,%
	citecolor=blue,%
	filecolor=black,%
	linkcolor=blue,%
	urlcolor=blue}

\begin{document}
	
\date{}

\maketitle

\singlespacing	


\begin{abstract}
	\noindent Despite the widespread implementation of conditional cash transfers in low- and middle-income countries, evidence on their long-term effects remains limited. This paper evaluates the impact of Ecuador's Human Development Grant on the formal labour market outcomes of children from eligible households. The programme---one of the first of its kind---was characterised by weak enforcement of its eligibility criteria. Using a regression discontinuity design, we find that the grant increased the probability of working in the formal sector by almost 13\% around 15 years after exposure, thereby helping to curb the intergenerational transmission of poverty. This positive effect is most likely to operate through human capital accumulation.\vskip 0.5em
	
	\vskip 0.5em
	
\end{abstract}

\onehalfspacing

\section{Introduction}\label{Section 1}

In 1997, Santiago Levy, Mexico's deputy finance minister, sought to overhaul the government's anti-poverty approach. Existing programmes provided assistance to the poor mainly in the form of food subsidies. Levy argued that these programmes were ineffective and inefficient. Levy based his reasoning on a central tenet of Economics that states that direct cash transfers are more effective at improving  the welfare of the poor than subsidies for specific consumer goods. On this basis, Levy proposed the first conditional cash transfer (CCT) programme, \textit{Progresa}, later renamed \textit{Oportunidades}. The most salient feature of CCTs is that, to be eligible for direct cash payments, recipients must meet requirements related to children's school attendance, medical check-ups or similar obligations.

Since Mexico launched \textit{Progresa}, many countries at different levels of development across all five continents have introduced such programmes. CCTs have three objectives. In the short term, they seek to alleviate the financial constraints of the poor. In the medium term, their main objective is to improve human capital---health and education---among the poor. Finally, they aim to reduce poverty in the long run by improving labour market participation and earnings. Therefore, the most distinctive long-term feature of CCTs is their objective of improving the economic outcomes of the next generation, thereby helping to prevent the transmission of poverty to the next cohort.

Paradoxically, while numerous studies highlight the short-term benefits of CCTs, particularly for school enrolment, evidence on their long-term impacts remains limited and inconclusive. Furthermore, some voices have warned against the common practice---and the potential bias associated with it---of rarely conducting long-term follow-ups on social interventions when short-run effects are small \parencite{leight2022}, as has sometimes been the case with CCT programmes. 

This paper examines the impact of one of the pioneering programmes of its kind, the Human Development Grant (HDG) launched in 2003 in Ecuador---where enforcement of the grant's eligibility criteria has historically been very weak---, on the future labour market outcomes of young adults who were eligible for the programme when they were children. Combining information from poverty censuses (the specific databases used to administer the programme) and Social Security records in a regression discontinuity design (RDD), we estimate the local intention-to-treat (ITT) effect of the HDG on the employment probability of employment in the formal economy in 2024 by eligible individuals in 2008/2009. Therefore, we assess the impact of being a potential recipient of the programme approximately 15 years after of eventual initial eligibility.

Overall, our findings suggest that the HDG fulfils its long-term objectives. A child who was eligible for the HDG in 2008/2009 was 3.8 percentage points more likely to work in the formal labour market roughly 15 years later than an ineligible child. Such a figure represents a relative increase of around 13\%. We argue that, for several reasons, this must be a lower bound of the actual impact of the grant. Our results are robust to an extensive battery of robustness checks. Based on the previous literature on the grant's short-term effects and our additional analyses, we suggest that human capital accumulation (mainly education) is likely to be the main driver of these results.

We contribute to the literature in two ways. First and most importantly, our work adds to the limited evidence on the long-term impact of CCTs \parencite{molina2019}. Second, given the weak enforcement of the requirements for receiving the subsidy, it provides additional evidence to inform the debate on the relevance of conditionality in such programmes \parencite{baird2013,baird2014}.  

The rest of the paper unfolds as follows. Section~\ref{Section 2} discusses the existing literature and frames our work within it. Section~\ref{Section 3} outlines our research design, including the institutional setting, data and empirical strategy. Section~\ref{Section 4} presents the results of our analysis. Section~\ref{Section 5} summarises and discusses the main implications of the research.

\FloatBarrier
\section{Background and related literature}\label{Section 2}

Conditional cash transfers have been the spearhead of anti-poverty policies in Latin America and the Caribbean and other low- and middle-income regions over the last two decades. After the pioneering Mexican and Brazilian experiences, the HDG, in its current version, was rolled out as one of the first programmes of this kind operating in the hemisphere \parencite{cecchini2011,rawlings2005b,villatoro2005}. Similar to most programmes of this kind, in addition to targeting socially disadvantaged households, a monthly cash payment is conditional on families meeting certain criteria related to children's school enrolment and attendance at medical check-ups. Nevertheless, as we explain in the next section, an idiosyncratic feature of the Ecuadorian programme is that the enforcement of those requirements has been, at best, quite weak. 

These programmes aim to improve the lives of households experiencing deprivation and to reduce future poverty. The conditions attached to the benefits are intended to encourage human capital accumulation—particularly among children—to prevent the intergenerational transmission of poverty \parencite{cecchini2011,fiszbein2009}. According to \textcite{fiszbein2009}, the rationale for imposing eligibility conditions rests on two main considerations. The first is the assumption that parental investment in children's human capital would otherwise be suboptimal in the absence of such requirements, mainly because of positive externalities related to education and health, information problems (e.g., about the returns to human capital), principal--agent problems (incomplete altruism or even conflicts between the two parents), or behavioural factors (excessively high discount rates on the part of parents). Political economy considerations are also relevant. Requiring poor households to meet certain criteria to access targeted benefits is intended to foster greater acceptance of such programmes among taxpayers than would be the case under unconditional transfers.

The literature that assesses the performance of such programmes is substantial \parentext{see, among many others, \textcite{baird2014} and \textcite{parker2007}}. The case of the Ecuadorian programme has received much attention from the research community. In addition to establishing its relevance for poverty alleviation \parencite{fiszbein2009,ordonez2015,wb2018}, this body of work has highlighted the positive impact of the programme on school enrolment—see \textcite{ponce2023} for a survey—and a reduction in child labour \parencite{martinez2012,edmonds2012,schultz2004}. Overall, the grant does not appear to have materially affected child development \parencite{paxson2010,ponce2010,fernald2011} or adult labour supply \parencite{bosch2019}, with mixed results for health \parencite{fernald2011,moncayo2019}.

According to \posscite{barrientos2012} discussion of the overall benefits of social protection, cash transfers of this kind can curb the intergenerational transmission of poverty through several channels. By alleviating credit constraints, they can contribute to human capital formation \parencite{baird2013,baird2014,parker2007}. Furthermore, they may foster investment in durable assets that generate future income streams \parencite{martinez2005,maluccio2010,gertler2012,blattman2020,gelders2019}. Relatedly, in some cases—mainly depending on the recipient (e.g., if only mothers receive the grant)—social protection benefits can alter household resource allocation such that families spend more on goods and services that specifically advance children’s interests \parencite{attanasio2010,attanasio2014a,angelucci2013,bergolo2018,macours2012,schady2008b}.

The number of studies on the long-term impact of these programmes is quite limited \parencite{molina2019} but it offers some grounds for optimism. Recent works that take advantage of experimental or quasi-experimental designs find that child eligibility for a conditional cash transfer improves educational attainment and labour market outcomes at a later age in Mexico \parencite{kugler2018,parker2023,araujo2021}, Brazil \parencite{laguinge2024}, Honduras \parencite{molina2020}, Nicaragua \parencite{barham2024} and Colombia \parencite{garcia2012}. Furthermore, this literature suggests additional benefits in the latter two countries (e.g., positive effects on cognitive development and health behaviours) \parencite{barham2013,garcia2012}.

Three of these works \parencite{barham2024,kugler2018,garcia2012} specifically refer to the impact on labour market formalisation, the main outcome on which we provide evidence. Moreover, the results of these studies are not entirely conclusive. In some studies, the results are only significant for certain population groups and their findings are sometimes contradictory. For instance, \textcite{barham2024} report a positive impact only for men, while other studies find the opposite---a significant effect only for women \parencite{parker2023,garcia2012,araujo2018}---or fail to find any significant impact on labour market outcomes \parencite{filmer2014}.

Using data from a randomised controlled trial and an RDD, \textcite{araujo2018} study the impact of the Ecuadorian HDG on test scores, educational attainment and work outcomes at a horizon of approximately 10 years among children who lived in households eligible for the transfer. They report much more nuanced results than those for other Latin American countries. Their findings indicate that early-childhood exposure did not improve test scores 10 years later, a positive effect on secondary school completion and a positive impact, albeit not robust, on female employment. A related work \parencite{mideros2021} evaluates social mobility in terms of a composite welfare index (based on a set of family assets and characteristics) among Ecuadorian households, leveraging the administrative registry used for targeting the grant (which, by design, suffers from high attrition rates, well above 50\%). Using a difference-in-differences strategy, the authors conclude that HDG eligibility enhances absolute and relative social mobility. Note that this study tracks changes in households' own socio-economic status over time, rather than assessing whether the grant improves the next generation’s well-being in adulthood.

Our work seek to contribute to this still-limited body of literature by providing additional evidence on the Ecuadorian case. Specifically, we focus on the likelihood of working in the formal economy in February 2024, among individuals who were eligible for the HDG in 2008/2009.

It is worth highlighting the substantial differences between the second analysis carried out by \textcite{araujo2018}---based on an RDD and centred on educational attainment and labour market outcomes---and ours, which together motivate this research. First, while \textcite{araujo2018} centre their analysis on the 19--25 age group, our research focuses on individuals aged 26--27, who are much more likely to have completed their transition from school to work (and, therefore, the risks associated with underestimating the programme's positive effects on employability are lower). For example, in the first quarter of 2024, labour-force participation was roughly 62\% among those aged 19--25 and 75\% among those aged 26--27, whereas the employment rate was 56\% and 68\% for these two groups, respectively \parencite{inec2025d}. In fact, \textcite{molina2019} argue that the lack of impact identified in pioneering studies of the long-term effects of CCTs may be due to the focus on individuals who were too young.

Second, the focus of these authors is on a different outcome (the probability of employment at a given point in time, in 2008/2009) from that considered here. In this regard, given Ecuador's low unemployment rate of 3.5\% in 2024 \parencite{ilo2025}, it is reasonable to infer that any improvements in labour market outcomes will be in terms of job quality rather than quantity, primarily via a reduction in informality, which affected almost 70\% of workers in that year.

Third, the identification strategy in \textcite{araujo2018} relies on eligibility for the HDG according to a poverty census from 2002/2003 (and the authors centre on outcomes in 2008/2009). We concentrate on a different time frame (from 2008/2009 to roughly 15 years later), making use of the same type of cadastre from 2008/2009, which aimed not only to update the registry of potential beneficiaries but also to significantly improve the targeting of the grant \parencite{fabara2009, mies2019}. Such changes could also enhance the performance of the HDG. Moreover, the HDG became more generous over time. The HDG amount was set at \SI{15}[US\$]{} per month from 2003 to 2007, rose to \SI{30}[US\$]{} in 2008 and then to \SI{35}[US\$]{} between 2009 and 2011. Since 2012 it has stood at \SI{50}[US\$]{} and since 2018 it has varied according to the number of children, reaching up to \SI{150}[US\$]{} \parencite{eclac2025}.

Last but not least, their analysis is based on the merging of two poverty censuses using the identity card number (\textit{c{\'e}dula de identidad}) of generally an adult woman (typically the mother of the children who made the household eligible to receive the HDG). In principle, this strategy has two limitations. In the first place, from 2002/2003 to 2008/2009, they are only able to track around 55\% of the households that were present in the 2008/2009 poverty census. This approach can result in non-random sample attrition. For families that believed they were suitable candidates for receiving welfare payments, ensuring their presence in the poverty census (which households could voluntarily participate in, even if they did not receive a visit from government interviewers) was clearly urgent. However, it is reasonable to suspect that those whose economic situation had improved over time had fewer incentives to remain present in the cadastre.\footnote{Therefore, in principle, if the programme has any effect on economic success, an exploration that relies on tracking individuals across different waves of the poverty census could underestimate the effect of eligibility on socio-economic outcomes (i.e., it would be less likely to observe those individuals exhibiting better economic performance over time).} In contrast, as explained in Subsection~\ref{Subsection 3.2}, in our analysis, we are able to recover the identity card number of more than 80\% of the children whom we aim to follow in 2008/2009, who can potentially appear in (and be linked to) Social Security records around 15 years later.

Secondly, \textcite{araujo2018} are able to examine only the outcomes of young adults who continued to live in the household in which they were raised (and not those who formed their own household or moved elsewhere). It is unlikely that the outcomes of those young adults aged 19--25 who are still living with their mothers are representative of the entire population of interest. They are likely to have had much lower labour market attainment than those who had already left the family home.

Our study also contributes to the existing literature on the relevance of conditionality associated with these types of benefits \parencite{baird2013,baird2014}. As we explain in Section~\ref{Section 3}, the enforcement of conditionality for the Ecuadorian HDG is weak, in contrast to other programmes of this kind in the region. Although national authorities have always publicly announced the requirements regarding school enrolment and medical check-ups, they have not monitored compliance. Consequently, the HDG cannot be described as a strict \textit{conditional} transfer. Some authors argue that the programme sits somewhere between a conditional and an unconditional transfer \parencite{schady2008a}. 

When assessing children's future performance in the formal labour market as young adults, we think it is useful to first describe the characteristics of this in Ecuador. The definition of formality used here follows the \enquote{legalistic} or \enquote{social protection} approach \parencite{gasparini2009}. Formal workers are those affiliated with Social Security and consequently enjoy entitlements to certain social benefits, such as contributory old-age, survivors' and disability pensions, maternity and sickness benefits, unemployment insurance and health care \parencite{ssa2020}. Informality is a multidimensional phenomenon associated with a range of explanatory factors (economic structure, lack of law enforcement, poor public services and burdensome regulatory frameworks). Whether participation in the informal economy is a voluntary decision is also a subject of debate \parencite{loayza2009,maloney2004,cimoli2006,portes1993,laporta2014,biles2009}. 

Informality in the Ecuadorian labour market is pervasive and almost endemic. This segment accounted for 72\% and 69\% of total employment in 2008 and 2024, respectively \parencite{ilo2025}. Regardless of the factors behind the existence of this sector, informal workers in Ecuador on average have much worse social outcomes than those in the formal labour market, such as lower earnings, higher poverty rates and worse future career prospects \parencite{canelas2019,matano2020,maurizio2019,maurizio2021,maurizio2023}. In addition, available evidence suggests that informality may also result in negative externalities, too. A larger informal sector not only results in higher inequality but also negatively affects tax collection \parencite{boitano2019}, the health status of the population \parencite{utzet2021}, pension coverage \parencite{daude2015} and levels of political participation \parencite{baker2022}. Therefore, even though working in the formal sector in Ecuador is not the only indicator of a programme's long-term success, it is undoubtedly a positive socio-economic outcome that is of clear interest to policymakers. 

\FloatBarrier
\section{Research design}\label{Section 3}
\subsection{Institutional setting}\label{Subsection 3.1}

The HDG was introduced in Ecuador in 2003 as a reformulation of the Solidarity Grant, an earlier and poorly targeted social benefit introduced in 1998 to provide a safety net for socially disadvantaged families. This grant aimed to compensate these households for the removal of gas, petrol and electricity subsidies, part of the liberalisation and adjustment policies adopted in the late 1990s. The HDG programme sought to alleviate poverty in the short term and to foster human capital formation in order to prevent intergenerational transmission of poverty. In particular, it targeted vulnerable households with at least one child younger than 16 years under the rules in force during our study period.

The 2003 redesign aimed to improve targeting. In particular, the Ecuadorian government---with technical assistance from local universities---created an ad hoc poverty census to better identify the most vulnerable population (in Spanish, \textit{Sistema de Identificaci{\'o}n y Selecci{\'o}n de Beneficiarios de Programas Sociales} [SELBEN]). Data collection on households’ socio-economic and demographic characteristics relied on home visits, public announcements and a voluntary enrolment mechanism through which households could request an eligibility assessment.
The national authorities initially developed an eligibility index (from 0 to 100, from the lowest to the highest well-being level) based on a principal component analysis of 27 household variables. Households in the first two quintiles of this index with children below 16 years old were eligible for the grant.

In principle, the programme imposed two conditions on grant recipients. The first requirement related to education: children aged six to 16 had to be enrolled in school and attend classes regularly (compliance was defined as having no more than four unjustified absences per two-month school period). The second condition was that children  up to five years old and pregnant women in beneficiary households were required to attend at least one preventive health visit every two months at a public health facility. Although the authorities envisaged sanctions for non-compliance (\SI{6}[\$]{} per month) and introduced a standard monitoring card as a verification instrument, the programme---unlike other CCTs operating in Latin America and the Caribbean---did not strictly enforce conditionality. Consequently, the Ecuadorian authorities did not suspend benefits if families failed to comply with the requirements, meaning that the transfer was unconditional in practice. Although the Ecuadorian government did not verify compliance, families did commit in writing to satisfying the conditions, and the authorities always publicly emphasised the need to meet the stated requirements. These conditions, which were slightly revised in 2013, were those in force for the individuals who were potentially eligible for the transfer and are considered in this study.

Abstracting from the non-enforcement of conditionality, loss of eligibility (due to changes in the presence of children or in the SELBEN index) implied benefit withdrawal. In practice, government employees regularly visited households to update the poverty census. This process may have resulted in the suspension of the HDG. Unfortunately, information on how often this occurred was scarce. For instance, the government maintained the grant as long as households had children under 18 years old. Loss of eligibility under this criterion should have been easy to monitor, but updates to eligibility status could take months or years to be implemented in practice. In addition, Ecuadorian authorities tended to withdraw the grants from large numbers of households at once rather than continuously on a case-by-case basis.

At the outset of the programme, the government committed to renewing the registry approximately every five years and, to this end, set up the so-called Social Registry 2008/2009. Administered by the Social Registry Unit, a public agency under the auspices of the Ministry of Economic and Social Inclusion of Ecuador, the Social Registry operated similarly to the SELBEN and aimed to improving the targeting of the HDG. Since the 2008/2009 wave, the National Institute of Statistics and Censuses of Ecuador has been responsible for data collection. Using non-linear component analysis and 30 household variables, the Social Registry Unit developed a Social Registry Index, rescaled from 0 to 100, to determine eligibility for different social benefits. The government set the cut-off point at 36.5987 points. Along with the rules related to the presence of children under 16 years old, this criterion determined eligibility for the grant from August 2009 to August 2014, when the new Social Registry 2013/2014 came into force. Following the 2008/2009 Social Registry update, benefits were withdrawn from some households and extended to newly eligible ones. This transition took place between August and December 2009 \parencite[][Figure 10.3]{araujo2018}. The process meant that HDG payments ceased for more than \num{200000} households \parencite{buser2017}.

On top of being the flagship social programme of successive Ecuadorian governments for more than two decades, the HDG has also become one of the most important benefits of its kind in Latin America and the Caribbean. In 2022, it represented approximately 1\% of GDP and reached 7.6\% of the population (more than 12\% in 2011). The basic amount of the HDG was \SI{11.5}[\$]{} per month from 2003 to 2007 (and up to \SI{15}[\$]{} until 2006 and \SI{30}[\$]{} in 2007 for the 20\% poorest households), \SI{30}[\$]{} in 2008,  and \SI{35}[\$]{} in the period 2009--2011 and has been \SI{50}[\$]{} since 2012 \parencite{eclac2025}. The size of the transfer was considerable, bearing in mind that the average national per capita income and the average per capita income of the first two quintiles in 2008 (our departure point) were {\SI{166}[\$]{}} and {\SI{32}[\$]{}} per month, respectively \parencite{cedlas2025}. Whenever possible, the recipients are mothers, who can withdraw the benefit from private banks. More recently, eligible mothers have also been able to receive the payment directly into their bank account.

\FloatBarrier
\subsection{Data}\label{Subsection 3.2}

Our analysis uses data from three sources. The first is the Social Registry 2008/2009 \parencite{socialregistry2025}. As explained above, this is a cadastre administered by the Ministry of Economic and Social Inclusion that contains static socio-economic and demographic information on Ecuadorian households. It enables public institutions to determine eligibility for social benefits and to achieve appropriate targeting according to the procedure described above.

The Social Registry database includes the Social Registry Index (and the component variables needed for its calculation) which governs eligibility for the HDG. From August 2009 onwards, households were eligible if they scored fewer than 36.5987 points on the index. The Social Registry 2008/2009 index---hereafter referred to as the poverty index---is therefore the running variable in our analysis. Furthermore, we have access to administrative data on actual HDG receipt in 2010. We combine these data with information from the Social Registry 2013/2014 \parencite{socialregistry2025}, which contains children's educational outcomes at that time---specifically, whether each child had completed primary education by the survey date. Although this is not a primary outcome, it is relevant for our analysis, as it allows us to explore the potential mechanism through which the HDG operates in the long term.

We merge these data from the Social Registry with information from \textcite{socialsecurity2025}, our third data source, for February 2024. It provides data on monthly labour income reported to Social Security. This administrative registry allows us to track formal-sector labour market performance in that month for all individuals with a valid identity card number who were included in the Social Registry in 2008/2009. Our primary outcome is employment in the formal sector, which has implications beyond earned labour income, such as access to social benefits and services. We also analyse formal labour income (unconditional on employment) as a secondary outcome in our sensitivity analysis.

With these data, we are able to identify the HDG eligibility status of children in the Social Registry 2008/2009, along with their subsequent participation in the formal sector when they were 26 or 27 years old in February 2024. To ensure that the children included in the analysis were exposed to a minimum treatment intensity, we restrict the sample to those who were 13 years old or younger in January 2010, when the transition from the old to the new payment system was completed. This ensures that children below the poverty index threshold considered in the analysis were exposed to HDG eligibility for at least two years. Consequently, our sample comprises \num{59299} children aged 12 to 13 in January 2010, who were eligible for the HDG for roughly two to three years depending on their age, and who were 26 or 27 years old in February 2024.\footnote{If there were other children under the age of 16 in the household, even if not eligible, members of our sample could still benefit indirectly from the transfer. Therefore, our estimates should be interpreted as a lower bound of the HDG's impact.} We focus on young individuals living in households included in the Social Registry 2008/2009 whose poverty index ranged from 30.5 to 42.5 points. This corresponds to the sampling frame provided by the Ecuadorian National Institute of Statistics, which was responsible for granting access to the data. This restriction is not problematic, as our analysis relies on local rather than global polynomial regressions.

For individuals younger than those in our analysis, high labour market participation is not necessarily a positive socio-economic outcome. Early entry into employment may simply indicate that a young person has left the education system prematurely. Therefore, the school-to-work transition should not play a major role in our analysis. The standard age for completing higher education in Ecuador is 21. For instance, in the first quarter of 2024, nearly 68\% of individuals aged 26 or 27 were employed (the labour force participation rate was approximately 75\%), and only around 12\% were still in education (about half of whom were also working). By contrast, about 52\% of adults aged 18--25 were employed (and around 58\% were part of the labour force), while almost 31\% remained in education \parencite{inec2025d}.

The merging of all databases, carried out by the National Institute of Statistics and Censuses of Ecuador, was performed using the children's national identity card number. We assume that individuals with a valid identity card number who were not found in the Social Security records did not participate in formal labour activities---that is, they had no formal employment and zero formal labour income. Unfortunately, a substantial proportion of individuals (18.6\%) lack an identity card number. This may be due either to the number not having been recorded by the registry administrators or to the fact that many children did not possess a national identity card in 2008/2009, when it was not yet compulsory for minors.

Moreover, the share of children without a personal identity card number changes significantly at the cut-off point, being higher above (24.8\%) than below the threshold (13.5\%). This is due to two different reasons. First, the authorities placed much greater emphasis on collecting detailed information on individuals below the cut-off (the actual potential recipients of the benefit) than on those above the cut-off. The latter group received much less attention from the Ecuadorian authorities as they were not relevant to programme administration. Second, the grant administrators encouraged household members receiving the benefit to obtain a personal identity card if they did not have one. These two strategies, adopted by the programme administrators, have a clear implication: it is reasonable to assume monotonic selection, namely that moving from below to above the cut-off does not make any individual more likely to have a valid identity card number (and, therefore, to be observed). We discuss in detail the implications of this circumstance for our analysis in the next subsection and how this shapes our empirical strategy.

Lastly, the merging of the two Social Registries is not without problems. In addition to the issue of missing identity numbers mentioned above, the process is affected by substantial attrition: only about 50\% of children can be tracked from 2008/2009 to 2013/2014. As in the previous case, we discuss the consequences of this issue below.

Table~\ref{Table 1} displays summary statistics for the primary outcome of interest (formal employment, unconditional on labour market participation) in February 2024, the running variable and the covariates considered in the analysis. Our complete-case sample comprises \num{48157} individuals who fulfil the conditions set out above.

\begin{center}
	\begin{singlespace}
		\begin{table}[ht]
			\centering
			\begin{threeparttable}
				\centering
				\def\sym#1{\ifmmode^{#1}\else\(^{#1}\)\fi}
				\footnotesize
				\setlength\tabcolsep{5pt}
				\sisetup{
					table-number-alignment = center,
				}
				\begin{TableNotes}[flushleft]\setlength\labelsep{0pt}\footnotesize\justifying
					\item\textit{Note:} All variables refer to 2008/2009, with the exception of age (measured in January 2010).
					\item\textit{Source}: Authors' analysis from \textcite{socialregistry2025} and \textcite{socialsecurity2025}.
				\end{TableNotes}
				\begin{tabularx}{0.65\textwidth}{X *{2}{S[table-column-width=1.7cm]}}
					\caption{Descriptive statistics} \label{Table 1} \\		
					\toprule
					&\multicolumn{1}{c}{Mean}&\multicolumn{1}{c}{\makecell{Standard\\deviation}}\\
					\midrule
					Formally employed in February 2024&0.274&\\
Poverty index in 2008/2009&35.986&3.386\\
Female&0.489&\\
Ethnic minority&0.161&\\
Age&12.910&0.287\\
Female household head&0.400&\\
Household head's age&41.629&8.633\\
Household head married or in union&0.666&\\
Household head's years of education&7.015&3.328\\
Employed household head&0.910&\\
Household size&4.918&1.585\\
Rural area&0.267&\\ [1ex]
No. of observations&\multicolumn{1}{c}{\hspace{2mm}\num{48157}}&
\\
					\bottomrule
					\insertTableNotes
				\end{tabularx}
			\end{threeparttable}
		\end{table}
	\end{singlespace}
\end{center}

\FloatBarrier
\subsection{Empirical strategy}\label{Subsection 3.3}

We employ a sharp RDD to evaluate the long-run impact of the HDG. If selection were not an issue, under the standard RDD assumptions we could recover the local average ITT effect for the whole population covered by the cadastre using a research design that exploits the discontinuity in eligibility at the cut-off point in the Social Registry 2008/2009 index (36.5987). Given the non-negligible presence of children without a valid identity card number (and, therefore, without an observable labour market outcome) and---as we show below---the discontinuity in this proportion at the cut-off, we are unable to estimate such a parameter consistently. However, we can estimate the average local ITT effect for the observed population by means of a sharp RDD on the selected sample.

In the context of an RDD, selection is only problematic if it changes discontinuously at the cut-off. The main identifying assumption of this type of research design is that, in the absence of treatment, potential outcomes vary smoothly with the running variable. If the probability of being observed (i.e., having a valid identity number) is also a smooth function of the running variable, the set of observed individuals just to the left and right of the cut-off remains comparable, and the RDD estimator remains unbiased for the local average ITT effect among those individuals observed irrespective of their treatment status.

Nevertheless, if the probability of observation itself jumps at the threshold, the composition of observed individuals changes abruptly at the cut-off. In such a case, any observed discontinuity in the outcome may capture not only the causal effect of the treatment but also differences in sample composition arising from differential selection, thereby violating the continuity condition required for identification of the local ITT. Below the cut-off, we observe those individuals who would be observed under both the control and treatment regimes (\textit{always observed}) and those only observed under treatment (\textit{entrants}). Above the cut-off, we only observe the former group. This concern is particularly relevant if selection into the group with a valid identity card number is correlated with unobservables that also affect future labour market outcomes. For example, if the propensity to hold a valid identification number in childhood were positively correlated with the propensity to work in the formal labour market in adulthood, our estimates of the HDG ITT effect would be conservative (i.e., biased towards zero). In our case, as we show below, there exist a clear discontinuity in the selection probability at the cut-off.

According to \textcite{dong2019}, the local average ITT effect for the observed sample (which she terms the \textit{intensive margin}) is still an informative parameter for policy design and evaluation. The intensive-margin effect does not usually capture a causal impact at the individual level, and it is typically different from the local average ITT effect for the whole sample or for the always-observed individuals. Nevertheless, one can view the local average intensive-margin ITT effect as a causal parameter from a distributional point of view (similar to the distributional effects in the programme evaluation literature). This parameter can still be of interest if policymakers care about how the observed outcome changes with treatment. In our setting, the intensive margin measures how formal employment among those with a valid identity card number in childhood varies with HDG eligibility, regardless of any composition change.

Furthermore, our knowledge of the selection procedure allows us to further characterise the meaning of the intensive-margin local average ITT effect in our setting. First of all, for the reasons behind the higher proportion of children with valid identity card numbers below the cut-off, it is reasonable to assume monotonic selection, which implies the absence of \textit{quitters} (children with a valid identity card number only if non-eligible). Under the assumption that entrants are not better off than always observed under non-eligibility (which is reasonable as long as they proved to be hard to reach by authorities or required additional encouragement to get a valid identity card), the intensive margin represents a lower bound for the effect among the group composed of the always-observed and entrant individuals. If we add the (also very reasonable) assumption of a non-negative effect on \textit{never-observed} individuals (those without a valid identity card number irrespective of eligibility), the intensive margin also becomes a lower bound for the local average ITT effect for the whole population covered by the registry.\footnote{Under monotonic selection and bounded outcomes, \textcite{dong2019} derive sharp bounds for the effect on the always-observed by replacing entrants’ unobserved outcomes with their lower and upper limits. In practice, the width of these bounds depends on the size of the selection discontinuity, the precision of the estimated jumps in selection and in the outcome, and the effective sample size near the cut-off. In our case, although treatment induces a clear jump in the probability of observation, the magnitude of the discontinuity in the selection probability is sizeable, the discontinuity in the outcome is relatively small and imprecisely estimated, and the number of individuals close to the cut-off is relatively limited. As a consequence, the resulting bounds are uninformative. Furthermore, the group of always-observed children, although well defined at the individual level, is not of particular interest from a policy perspective. We therefore focus on the intensive-margin effect among observed individuals, which uses more than 80\% of the registry and remains meaningful for policy as long as it constitutes a lower bound for the local average ITT effect for both the whole population covered by the registry and the group composed of always-observed and entrant individuals.}

The local average intensive margin ITT effect can be formally defined as \parencite{dong2019}:
\begin{equation}
	\label{Equation 1}
	\E\left[Y\left(1\right) \,\middle|\, S(1) = 1,\, X = c\right] - \E\left[Y\left(0\right) \,\middle|\, S(0) = 0,\, X = c\right],
\end{equation}
where $Y(t)$ and $S(t)$ are the potential outcome (formal employment) and the potential sample selection under non-eligibility or eligibility ($t=0,1$), respectively; $X$ denotes the forcing variable (the poverty index) and $c$, the cut-off. Unlike the fuzzy case, this expression can be estimated via a sharp RDD on the observed sample.

All our analyses rely on non-parametric local polynomial estimation methods using a data-driven mean squared error (MSE) optimal bandwidth \parencite{calonico2019a,calonico2019b,cattaneo2020a,cattaneo2022} and assess its sensitivity in our robustness checks. Our rationale for choosing this method is that global high-order polynomials (i.e., a parametric approach) lead to noisy estimates, sensitivity to the polynomial degree and poor coverage of confidence intervals \parencite{gelman2019}. 

Our application of this method unfolds as follows \parencite{cattaneo2020a}. First, we select a polynomial of order $p$ and a kernel $K(\cdot)$. Standard practice is to choose $p = 1$ (local linear regression). We also employ $p = 2$ (local quadratic regression) as a robustness check. Second, we choose a bandwidth. We select a bandwidth $h$ that minimises the asymptotic MSE of the point estimator for our baseline analysis \parencite{calonico2019b}. We also consider, in our sensitivity analyses, an alternative optimal bandwidth that minimises the coverage error rate (CER) of the confidence interval.

Third, we choose a kernel function to weigh the observations in the interval of interest around the cut-off point ($c$). We select a triangular kernel, $K(u) = (1 - \lvert u \rvert)\mathbb{1}(\lvert u \rvert \leq 1)$, which assigns non-negative weights to each transformed observation (centred around the cut-off and then divided by the selected bandwidth) based on the distance between the observation's score and the cut-off. This kernel function is MSE-optimal.

Fourth, we estimate a weighted least squares regression, with weights given by the kernel function $K\left(\frac{X_i - c}{h}\right)$, of the outcome variable for individual $i$ at time $t$ ($Y_{i}$)---an indicator that takes the value one if the person is employed in the formal sector and zero otherwise---on an intercept, the treatment indicator $T_i$ (equal to one if the poverty index in 2008/2009 is less than or equal to the cut-off point and zero otherwise), the polynomial in the re-centred forcing variable ($X_i - c$), its interaction with the treatment indicator, and the set of control covariates of interest ($Z_i$). The set of control variables includes child's gender, child's age, child's ethnicity, household head's gender, household head's age, household head's age squared, household head's marital status, household head's head years of education, household head's employment status and area and province of residence. All covariates are measured in 2008/2009 (with the exception of age, measured in January 2010 for all children). 

Formally, we estimate the parameter capturing the local average intensive margin ITT effect ($\hat{\tau}$) on through the following equation:
\begin{equation}
	\label{Equation 2}
	\hat{Y}_{it} = \hat{\alpha} + \hat{\tau}T_i + \hat{\mu}_{+} (X_i - c) + \hat{\mu}_{-} (X_i - c) T_i + Z_i' \hat{\Theta}
\end{equation}
We assess the precision of this point estimate using robust bias-corrected confidence intervals (more conservative than conventional ones) \parencite{calonico2014b} and we cluster the standard errors at the household level \parentext{the level at which the forcing variable determining treatment varies \parencite{abadie2023}}.\footnote{In practice, one should report the conventional point estimate and assess statistical significance using the bias-corrected estimate and its robust standard error. For this reason, it is uninformative to report either the standard error of the conventional estimate or the robust standard error of the bias-corrected estimate. Accordingly, throughout we report the conventional point estimate together with the 90\% robust bias-corrected confidence interval for the bias-corrected estimate.} Within this framework, the inclusion of controls primarily aims to improve efficiency, rather than to correct unexpected imbalances (which would require invoking parametric assumptions on the regression functions to allow extrapolation or redefining the parameter of interest) \parencite{cattaneo2020a}.

Note that households below the relevant threshold not only received the grant when the new targeting rule came into force, but were also likely to continue receiving it until their eligibility status changed (and this was detected by the authorities) or until the Social Registry Unit updated the cadastre (for instance, if there were eligible siblings). Hence, our sample construction ensures that households below the cut-off likely benefited from the HDG for at least two years, unless they lost eligibility. According to the administrative payment records for 2010 (the only year for which such data are available), the programme's targeting performance---relative to the database information for households with an index value between 30.5 and 42.5 points---was excellent, with very low undercoverage (1.8\% of children in households below the threshold did not receive the grant) and leakage (only 7.9\% of those in families above the cut-off received payments).\footnote{These figures refer only to households included in our Social Registry sample, i.e. those with poverty-index values between 30.5 and 42.5 points. When considering the entire population theoretically eligible for the grant, take-up rates remain high (65\% in the two poorest quintiles in 2013), although travel costs, identity-related stigma and dissatisfaction with the government pose significant barriers to claiming the HDG \parencite{rinehart2017}.} Therefore, our estimates capture not only the intrinsically interesting ITT effect but also provide a reasonable lower bound for the local average treatment effect.\footnote{Although we observe receipt of the HDG in 2010, we do not use this information to estimate the local average treatment effect via a fuzzy RDD. Apart from the almost perfect compliance, there are two reasons for this choice. First, the discontinuity in the cut-off is very likely to be correlated with receipt of the HDG not only in 2010 but also in subsequent years, for which we lack administrative data. Consequently, we do not consider this strategy to offer a better understanding of the long-run effects of the grant. Second, in the presence of a discontinuity in the probability of selection (as in our case), inference becomes more complex and requires bootstrap procedures \parencite{dong2019}. In practice, this results in lower precision and precludes the use of robust bias-corrected inference, as recommended by \textcite{cattaneo2020a}.}

In order to formally evaluate the magnitude of the discontinuity in selection, we employ the specification embedded in Equation~\ref{Equation 2} (using the full sample) without covariates and with observability (i.e., having a valid identity card number) as the outcome. We find that crossing the threshold decreases the probability of having a valid identity card number by 11.7 percentage points (Figure~\ref{Figure A1}), which is sizeable.

The source of identification of the ITT effect in the observed sample is the reduction in households' likelihood of being paid the HDG upon crossing the eligibility threshold in the Social Registry Index 2008/2009. In other words, the policy application is as good as randomised in the neighbourhood of the cut-off provided the research design satisfies certain conditions (apart from monotonicity in selection) \parencite{dong2019}.

The first condition is that there must be no manipulation of the forcing variable by families (in the whole sample). In our setting, it is virtually impossible for families to alter their position relative to the threshold, since the Ecuadorian authorities determined the value of the threshold only after constructing the entire database.\footnote{First, the authorities collected all information on household characteristics through interviews. Second, they constructed a poverty index using a non-linear principal components analysis approach. Finally, they chose as the threshold the value that excluded the bottom 40\% of households in terms of socio-economic status. Thus, the information on households was collected prior to the determination of the cut-off, precluding any strategic behaviour by respondents.} In any case, we perform a manipulation test of the density discontinuity based on the local polynomial density methods proposed by \textcite{cattaneo2018,cattaneo2020b}. The results of this test do not allow us to reject the null hypothesis of no manipulation (Figure~\ref{Figure A2}). When looking at the complete-case sample, there is a clear discontinuity in the density at the cut-off point (see Figure~\ref{Figure A3}). It is important to note, however, that this latter finding should hardly be interpreted as evidence of manipulation (given the way the threshold was determined) and is better seen as a mechanical result of the selection process.

The second condition is that there must be no correlation between an observation being below the cut-off point and the factors affecting the labour market outcome. We assess whether there is any discontinuity in the average values of the observable covariates (measured almost 15 years prior to the realisation of the outcome) using the specification outlined above (but without covariate controls) on the whole sample. We reject the null hypothesis of continuity in only 4 out of 46 predetermined characteristics---less than 10\% of the cases---,so we cannot rule out that this result is due to chance (Table~\ref{Table A1}). Moreover, we report similar results for the selected sample (Table~\ref{Table A2}), which suggests that the resulting bias due to unobservables should be relatively modest.

Additionally, in order to assess the relevance and stability of our estimates, among our battery of robustness checks, we implement two different procedures that attempt to correct for selection and provide an estimate of the local average ITT effect for the whole population of the registry under the missing completely at random (MCAR) or missing at random assumption (MAR$|X$) (whereby the probability of an observation being missing depends only on the variables used for the correction). First, we rely on state-of-the-art multiple imputation (MI) methods of the missing outcome \parencite{vanbuuren2018}. Specifically, we model formal employment using a logistic regression as a function of a set of observed covariates (time-invariant or measured at baseline) that are likely to affect the outcome (formal employment in February 2024) or to be associated with non-response.\footnote{We consider the same covariates included in Equation~\ref{Equation 2}, plus the household head’s affiliation to Social Security, the number of children aged 15 or under in the household, the number of adults aged 65 and above in the household, type of dwelling, housing tenure, number of rooms, land ownership, ownership of draft animals, ownership of mobile phones, and an indicator for having an emigrant household member. We also include the poverty index, the household’s position relative to the cut-off (below or above), and interactions between that position and all of the aforementioned observable characteristics.} Using this method, we impute formal employment for 100\% of observations with a missing identity card number.

MI generates several plausible versions of the dataset and combines estimates across them using \posscite{rubin1987, rubin1996} rules. Apart from providing consistent estimates under MCAR or the MAR$|X$ assumptions, this approach also yields valid inference, as the combined variance reflects both within- and between-imputation uncertainty \parencite{little2019, murray2018}. Consequently, standard errors are asymptotically correct and confidence intervals achieve nominal coverage rates \parencite{vonhippel2020}. Since current software implementations do not accommodate the RD routines used in this paper within an MI framework, we combine the parameter estimates and variances obtained from each imputed dataset externally.

We set the number of imputations to 20. Based on our main analysis under our preferred specification (with extensive controls), the estimated fraction of missing information for our main parameter is 16.3\%, implying a relative efficiency of 99.3\%. The resulting loss of precision is therefore negligible. In line with the multiple-imputation literature, relative efficiencies above 99\% are generally considered fully adequate for valid inference \parencite{rubin1987, little2019, graham2007, vonhippel2020}. This number of imputations ensures stable standard errors and nominal coverage even with moderate fractions of missing information, while additional imputations yield minimal gains in precision.

Moreover, we present the results of our main analyses using inverse probability weighting (IPW). We first estimate the probability that an observation has a valid outcome as a function of the same covariates used for MI (allowing for side-specific effects of all covariates), and then use the inverse of this predicted probability as an observation-specific weight in the RD regressions. We construct stabilised weights by trimming the estimated propensities at the \nth{1} and \nth{99} percentiles and then normalise them to have mean one. Weight diagnostics suggest stable weights, with a low coefficient of variation (i.e., little dispersion of weights relative to their mean) and a high effective sample size relative to the nominal sample size (i.e., little loss of precision from weighting).\footnote{Within the preferred local window around the cut-off, the fifth–to–ninety-fifth percentile ranges on each side overlap on $[0.672,,0.946]$, covering 56.8\% of observations. There is a pronounced right tail (the share with an estimated probability above 0.95 is 32.6\%) and essentially no mass near zero. Weight diagnostics indicate stable weights: the coefficient of variation is about 0.191 and the effective sample size is about 96.5\% of the nominal sample size.} Both IPW and MI rely on the MAR$|X$ assumption, under which they yield consistent estimates, although MI is typically more efficient as it also exploits information from incomplete cases \parencite{little2019,cole2008,seaman2013}.

While the MAR$|X$ assumption is untestable, if these correction procedures do not materially change the estimate, then selection on observables is unlikely to be very relevant, which may suggest that selection on unobservables is not very important either.

A common threat to identification in any sort of longitudinal study examining the impact of public policies on labour market outcomes is selective migration. While our dataset does not allow us to directly observe this phenomenon, we can discuss this issue and how it may potentially affect the interpretation of our results.

First, if the HDG has no effect on migration, and given that we do not reject continuity in observable characteristics around the cut-off, it is plausible to expect continuity in unobservable factors as well, including geographic mobility. Second, the use of Social Security records, which constitute a nationwide administrative register, allows us to adequately account for internal migration within the country. Although we cannot identify who has moved and where, employment spells remain recorded as long as individuals are affiliated with the national system.

Third, more importantly, potential bias could arise if HDG eligibility affected international migration, a highly relevant phenomenon for the Ecuadorian economy. The direction of this bias depends on whether the programme increased or reduced migration abroad. If the grant encouraged emigration, our RD estimates would understate the programme's impact, as emigrants cannot be linked to Social Security records and are therefore implicitly assumed to have zero employment. Conversely, if the transfer discouraged migration, our estimates would be biased upward. Although direct evidence for Ecuador is lacking, recent studies suggest that conditional cash transfers can increase geographic mobility \parencite{angelucci2015, araujo2021, barham2024, molina2020}, which would lead us to interpret our estimates as a lower bound of the true effect of the HDG.

Finally, given the importance of migration in Ecuador, it is worth addressing this issue more thoroughly, despite the limitations of our data. According to \textcite{olivie2008}, Ecuadorian emigrants who send remittances---the subgroup for which data are available---tend to be more educated and come from wealthier regions of the country. In contrast, the HDG targets households in the bottom two quintiles of the poverty index, so our cut-off separates the poorest 40\% of the registry (which, as a whole, is already less well off than the general population) from the rest from the rest. Migration is therefore likely to be considerably more prevalent among socio-economic groups above the threshold than among those we study. Since our identification strategy relies on non-parametric local polynomial methods estimated within a narrow window around the cut-off, selective migration should not pose a major concern for the validity of our results.

Regarding the problematic link between the 2008/2009 and 2013/2014 editions of the Social Registry, discussed above, it becomes obvious that the strong attrition affecting the database is likely to be non-random, as individuals may leave the poverty census when their economic situation improves. Consequently, the effect of the HDG on education could be underestimated. The problem is therefore more severe than in the case of the Social Security data (where missingness of the outcome affects fewer than 20\% of children), and we do not aim to correct it. Instead, we use this information solely for an auxiliary analysis exploring the potential mechanisms behind the impact of the HDG on labour market outcomes.

\FloatBarrier
\section{Results}\label{Section 4}

We present our estimation results in four steps. First, we discuss the impact of a household being below the index cut-off (i.e., the local average intensive margin ITT effect) on future labour market outcomes, presenting both a graphical illustration of the impact of the discontinuity and the econometric results. Second, we show the results of an extensive battery of robustness checks. Third, we look at whether the effects differ across different groups of children. Fourth, we discuss the potential mechanisms behind our main results based on additional analyses and references to previous literature on the effect of the HDG on earlier socio-economic outcomes. 

\FloatBarrier
\subsection{Main results}\label{Subsection 4.1}

\begin{figure}[htbp]
	\footnotesize
	\captionsetup{width=0.75\textwidth}
	\caption{Local average intensive margin ITT effects on labour market outcomes in 2024}
	\label{Figure 1}	
	\centering
	\noindent\makebox[\textwidth][c]{
		\noindent\begin{minipage}[c]{0.75\textwidth}
			\centering	
			\includegraphics[width=1\textwidth]{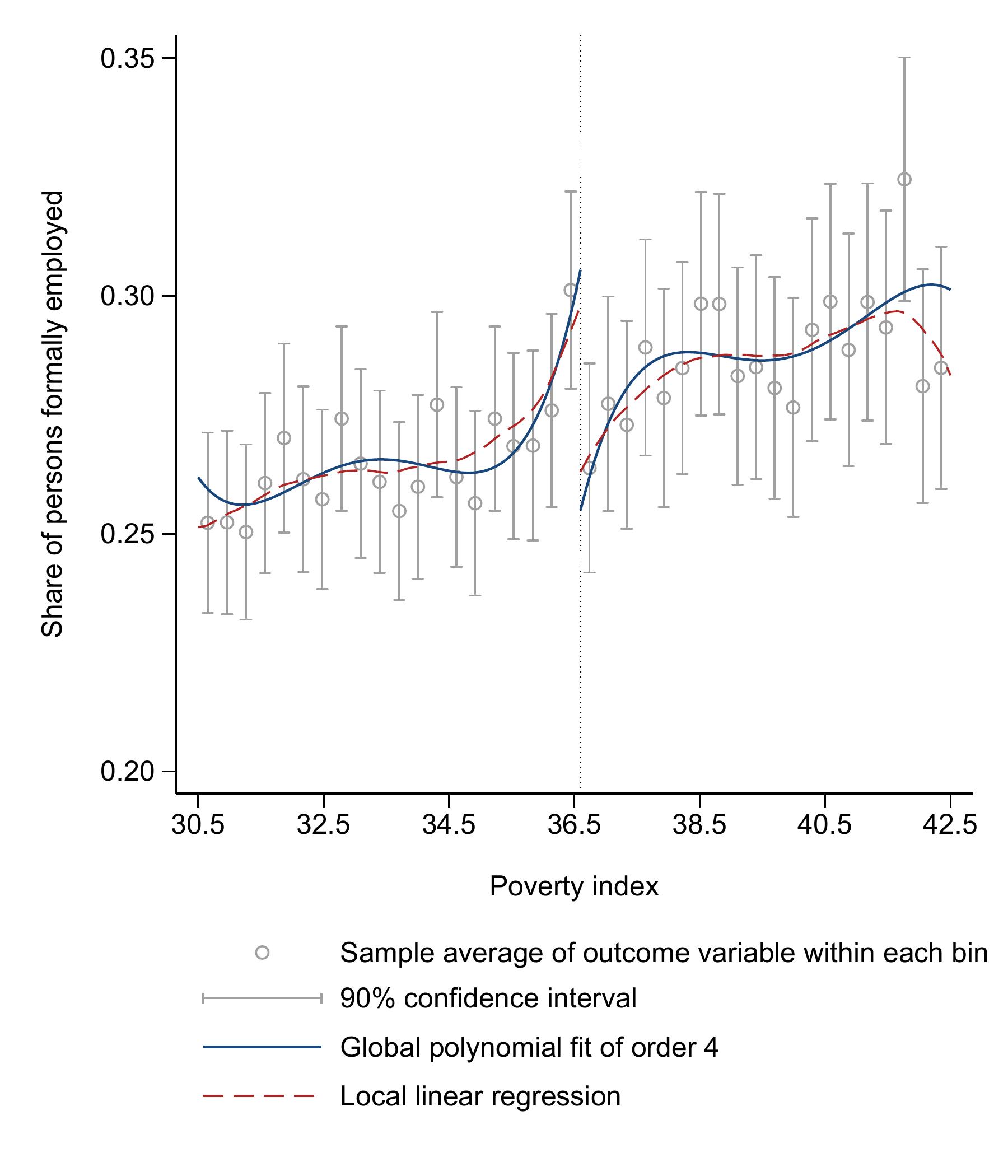} \\
			\justifying
			\noindent\textit{Notes}: The figure shows (i) the sample proportion of individuals employed in the formal sector within each bin and its 90\% confidence interval (grey circles); (ii) the fits of global fourth-order polynomial regressions estimated separately on each side of the cut-off (blue line); and (iii) the fits of local linear regressions estimated separately on each side of the cut-off using a triangular kernel and data-driven MSE-optimal bandwidths (red line).\\
			\noindent\textit{Source}: Authors' analysis from \textcite{socialregistry2025} and \textcite{socialsecurity2025}.
		\end{minipage}
	}
\end{figure}

Figure~\ref{Figure 1} illustrates the impact of HDG eligibility on the probability of employment in the formal sector when the individual is older than 25 years. Following \textcite{cattaneo2020a}, the graph presents local sample means, a global fourth-order polynomial fit and a local linear regression fit. Both parametric and non-parametric methods suggest that the HDG has a positive impact on the likelihood of employment in the formal sector. As discussed above, we favour the latter method (based on local linear regressions) for the remainder of our analysis, due to its advantages it has over the former, as highlighted in the literature.

\FloatBarrier
We report the main results of our econometric analysis in Table~\ref{Table 2}. We present three specifications: one without controls (I), one including province fixed effects (II), and one that adds a set of observable household socio-economic characteristics (III). As expected, the results are remarkably stable across specifications, since the inclusion of covariates is intended to increase efficiency. The econometric evidence confirms the message from the graphical exploration above: eligibility for the grant substantially improves young adults’ chances of formal employment. Focusing on column (III), being eligible for the HDG in 2010 for at least two years increases the probability of formal employment by 3.7 percentage points. This effect is sizeable: it represents a 12.7\% relative increase in the probability of employment in the formal sector in our sample.

\begin{singlespace}
	\begin{table}[htbp]
		\begin{ThreePartTable}
			\def\sym#1{\ifmmode^{#1}\else\(^{#1}\)\fi}
			\footnotesize
			\setlength\tabcolsep{5pt}
			\sisetup{
				table-number-alignment = center,
				range-phrase = {, }, 
			}
			\begin{TableNotes}[flushleft]\setlength\labelsep{0pt}\footnotesize\justifying
				\item\textit{Notes}: \sym{***} significant at 1\% level; \sym{**} significant at 5\% level; \sym{*} significant at 10\% level. Statistical significance is based on robust bias-corrected inference. Standard errors are clustered at the household level. 90\% confidence intervals based on robust bias-corrected inference are in parentheses. The table shows the results of local linear regressions with formal employment as the outcome estimated using a triangular kernel and data-driven MSE-optimal bandwidths. The control variables included in the third columns are child's gender, child's age, child's ethnicity, household head's gender, household head's ethnicity, household head's marital status, household head's educational attainment, household head's employment status, household size and area of residence (urban or rural).
				\item\textit{Source}: Authors' analysis from \textcite{socialregistry2025} and \textcite{socialsecurity2025}.				
			\end{TableNotes}
			\begin{tabularx}{0.9\textwidth}{X *{3}{S[table-column-width=2.5cm]}}
				\caption{Local average intensive margin ITT effects on formal employment in 2024} \label{Table 2}\\
				\toprule
				&\multicolumn{1}{c}{(I)}&\multicolumn{1}{c}{(II)}&\multicolumn{1}{c}{(III)}\\
				\midrule
				ITT effect&0.035\sym{**}&0.036\sym{**}&0.038\sym{**}\\
&[\numrange{0.010}{0.072}]&[\numrange{0.012}{0.074}]&[\numrange{0.013}{0.075}]\\ [1ex]
No. of observations &\multicolumn{1}{c}{\hspace{1.5mm}\num{48157}}&\multicolumn{1}{c}{\hspace{1.5mm}\num{48157}}&\multicolumn{1}{c}{\hspace{1.5mm}\num{48157}}\\
No. of observations effectively used  &\multicolumn{1}{c}{\hspace{1.5mm}\num{14590}}&\multicolumn{1}{c}{\hspace{1.5mm}\num{14023}}&\multicolumn{1}{c}{\hspace{1.5mm}\num{13922}}\\
Mean of dependent variable&0.274&0.274&0.274\\ [1ex]
Province fixed effects&&\multicolumn{1}{c}{\checkmark}&\multicolumn{1}{c}{\checkmark}\\
Control variables&&&\multicolumn{1}{c}{\checkmark}
\\
				\bottomrule	
				\insertTableNotes
			\end{tabularx}
		\end{ThreePartTable}
	\end{table}
\end{singlespace}

\FloatBarrier
\subsection{Robustness checks}\label{Subsection 4.2}

In this subsection, we carry out an extensive battery of robustness checks to assess the stability of our main results. First, we explore the impact of HDG eligibility on a secondary but closely related outcome: formal labour income (unconditional on labour market participation). Second, we present the results based on the sample with imputed outcomes for individuals without a valid identity card number according to the procedure outlined in Subsection~\ref{Subsection 3.3}. Third, we carry out the analysis using IPW, a different way of mitigating selection on observables. Fourth, we evaluate the impact of selecting a CER-optimal bandwidth. Fifth, we redo the main analysis using alternative kernels (the Epanechnikov and the uniform kernel). Sixth, we re-estimate the ITT effect on the selected sample using a local quadratic regression instead of a linear one. Seventh, we perform the analysis reducing and increasing the bandwidth by 20\%. Lastly, we repeat the analysis by excluding observations with a poverty index within 5\% of the bandwidth around the cut-off (a \enquote{donut hole} RDD).\footnote{This approach means excluding 675 observations of the total sample and reducing the effective sample used for estimation by \num{2195} observations, i.e., more than 15\% relative to the main specification.}

The results of these sensitivity analyses, presented in Tables~\ref{Table A3} and \ref{Table A4}, are all reassuring, allowing us to draw some meaningful conclusions. First, the impact on formal labour income is significant and positive, amounting to \SI{29.5}[US\$]{}, almost 18\% of the mean. The results for this secondary outcome and formal employment remain significant after adjusting for multiple testing \parencite{clarke2020}. Second, both MI and IPW yield significant, positive results that are not statistically different from our main intensive-margin estimates. Therefore, we cannot rule out that differences in point estimates are due to sampling variability. The fact that the results do not change much when accounting for potential selection on observables indicates that such selection is not very relevant and may also suggest that selection on unobservables is limited. Third, our results hold under the consideration of a CER-optimal bandwidth, alternative kernels, a quadratic polynomial regression, a narrower or wider bandwidth, or the exclusion of observations close to the cut-off. Finally, the two falsification tests suggest the absence of relevant discontinuities at the two placebo points, which reaffirms our confidence in our identification strategy.

\FloatBarrier
\subsection{Heterogeneity in the effects of the programme}\label{Subsection 4.3}

To study the impact of the HDG on different groups of individuals, we conduct a heterogeneity analysis by gender, ethnicity and area of residence (urban or rural) in 2008/2009, using the single-equation approach proposed by \textcite{calonico2025a, calonico2025b}. Table~\ref{Table 3} presents the results of this exercise. First, the estimated effect is statistically significant for men but not for women. However, although the point estimates differ markedly, the difference in coefficients across groups is not statistically significant, suggesting a lack of statistical power to detect heterogeneous impacts, as indicated by the wide confidence intervals for each group.

Second, a similar pattern emerges for ethnicity, where only the estimated coefficient for whites and mestizos is statistically different from zero, and we cannot reject the null hypothesis of homogeneous effects across ethnic groups. Lastly, regarding area of residence in 2008/2009, we can reject the null only for children living in urban households. Again, we cannot rule out that the effects are equal across groups, although in this case the point estimates are remarkably similar. This pattern suggests that the rural sample does not provide sufficient statistical power to detect the impact of HDG eligibility.

\FloatBarrier
\begin{landscape}
	\begin{singlespace}
			\begin{table}[p]
					\begin{ThreePartTable}
							\centering
							\def\sym#1{\ifmmode^{#1}\else\(^{#1}\)\fi}
							\scriptsize
							\setlength\tabcolsep{5pt}
							\sisetup{
									table-number-alignment = center,
									range-phrase = {, },
								}
							\begin{TableNotes}[flushleft]\setlength\labelsep{0pt}\footnotesize\justifying
									\item\textit{Notes}: \sym{***} significant at 1\% level; \sym{**} significant at 5\% level; \sym{*} significant at 10\% level. Statistical significance is based on robust bias- corrected inference. Standard errors are clustered at the household level. 90\% confidence intervals based on robust bias-corrected inference are in parentheses. The table shows the results of local linear regressions with formal employment as the outcome, estimated using a triangular kernel and data-driven MSE-optimal bandwidths. The control variables included in the third columns are child gender, child age, child ethnicity, household head' sex, household head's age, household head's age squared, household head's marital status, household head's educational attainment, household size and area of residence (urban or rural). The specification for each group excludes the covariate that defines that population segment (e.g., the analysis for men and women exclude the gender covariate). 
									\item\textit{Source}: Authors' analysis from \textcite{socialregistry2025} and \textcite{socialsecurity2025}.
								\end{TableNotes}
							\begin{tabularx}{\linewidth}{X *{9}{S[table-column-width=1.8cm]}}
									\caption{Heterogeneous local average intensive ITT effects on formal employment in 2024} \label{Table 3} \\		
									\toprule
									&\multicolumn{1}{c}{(I)}&\multicolumn{1}{c}{(II)}&\multicolumn{1}{c}{(III)}&\multicolumn{1}{c}{(IV)}&\multicolumn{1}{c}{(V)}&\multicolumn{1}{c}{(VI)}&\multicolumn{1}{c}{(VII)}&\multicolumn{1}{c}{(VIII)}&\multicolumn{1}{c}{(IX)}\\ [6pt]
									&\multicolumn{3}{c}{Child gender}&\multicolumn{3}{c}{Child ethnicity}&\multicolumn{3}{c}{\makecell{Area of residence in 2008/2009}}\\  [6pt]	
									&\multicolumn{1}{c}{\makecell{Males}}&
									\multicolumn{1}{c}{\makecell{Females}}&\multicolumn{1}{c}{\makecell{Difference\\$\text{(I)}-\text{(II)}$}}&
									\multicolumn{1}{c}{\makecell{Whites and\\mestizos}}&					
									\multicolumn{1}{c}{\makecell{Minorities}}&\multicolumn{1}{c}{\makecell{Difference\\$\text{(IV)}-\text{(V)}$}}&
									\multicolumn{1}{c}{\makecell{Urban}}&
									\multicolumn{1}{c}{\makecell{Rural}}&\multicolumn{1}{c}{\makecell{Difference\\$\text{(VII)}-\text{(VIII)}$}}\\[6pt]
									\midrule
									ITT effect&0.065\sym{**}&0.017&0.048&0.046\sym{***}&0.010&0.036&0.045\sym{***}&0.045&0.001\\
&[\numrange{0.008}{0.159}]&[\numrange{-0.013}{0.114}]&[\numrange{-0.050}{0.116}]&[\numrange{0.018}{0.126}]&[\numrange{-0.092}{0.151}]&[\numrange{-0.069}{0.154}]&[\numrange{0.032}{0.151}]&[\numrange{-0.073}{0.104}]&[\numrange{-0.013}{0.166}]\\ [1ex]
No. of observations &\multicolumn{1}{c}{\hspace{1.5mm}\num{24592}}&\multicolumn{1}{c}{\hspace{1.5mm}\num{23565}}&&\multicolumn{1}{c}{\hspace{1.5mm}\num{40387}}&\multicolumn{1}{c}{\hspace{3mm}\num{7770}}&&\multicolumn{1}{c}{\hspace{1.5mm}\num{35296}}&\multicolumn{1}{c}{\hspace{1.5mm}\num{12861}}&\\
No. of observations effectively used&\multicolumn{1}{c}{\hspace{3mm}\num{6462}}&\multicolumn{1}{c}{\hspace{3mm}\num{5993}}&&\multicolumn{1}{c}{\hspace{1.5mm}\num{10681}}&\multicolumn{1}{c}{\hspace{3mm}\num{1962}}&&\multicolumn{1}{c}{\hspace{3mm}\num{8576}}&\multicolumn{1}{c}{\hspace{3mm}\num{3964}}&\\
Mean of dependent variable&0.336&0.210&&0.279&0.251&&0.271&0.283\\  [1ex]
Province fixed effects&\multicolumn{1}{c}{\checkmark}&\multicolumn{1}{c}{\checkmark}&&\multicolumn{1}{c}{\checkmark}&\multicolumn{1}{c}{\checkmark}&&\multicolumn{1}{c}{\checkmark}&\multicolumn{1}{c}{\checkmark}\\
Control variables&\multicolumn{1}{c}{\checkmark}&\multicolumn{1}{c}{\checkmark}&&\multicolumn{1}{c}{\checkmark}&\multicolumn{1}{c}{\checkmark}&&\multicolumn{1}{c}{\checkmark}&\multicolumn{1}{c}{\checkmark}
\\
									\bottomrule
									\insertTableNotes
								\end{tabularx}
						\end{ThreePartTable}
				\end{table}
		\end{singlespace}
\end{landscape}

\FloatBarrier
\subsection{Mechanisms}\label{Subsection 4.4}

As discussed in Section~\ref{Section 2}, human capital formation is the most obvious channel through which the HDG may affect long-term socio-economic outcomes. To test the plausibility and relevance of this channel, we track the schooling performance of a subset of our sample in the Social Registry 2013/2014, when these children were aged 15--18 years. Specifically, we examine the local average intensive-margin ITT effects of baseline HDG eligibility on the probability of having completed primary education by the interview date, as recorded in the Social Registry 2013/2014. Using the same specifications as in our main results, we find that HDG eligibility increased the likelihood of completing this schooling level by 1.7 percentage points (Table~\ref{Table A5}). Given the selective attrition of our sample—probably reflecting exit from the cadastre among families with better economic outcomes—this estimate should be viewed as conservative. This result is consistent with previous evidence on the grant’s impact on educational outcomes. In particular, the HDG increases primary school enrolment by roughly 10 percentage points among children aged 6--17 \parencite{schady2008a}, and it raises the probability that young adults aged 19--25 have completed secondary education by about 1.5 percentage points. In the same vein, losing the right to receive the grant results in a substantial drop in enrolment rates in primary, secondary and higher education \parencite{merino2025}.

While we cannot provide additional direct evidence on other channels from our database (as Social Security records only contain information on formal labour market outcomes), we can interpret our results in light of prior evidence on other pathways. With respect to health, a recent non-experimental study using county-level panel data suggests that the expansion of the HDG led to a substantial reduction in under-five mortality---particularly from poverty-related causes, notably malnutrition, diarrhoeal diseases and lower respiratory tract infections---\parencite{moncayo2019}.\footnote{Additional evidence on positive effects of the HDG on health outcomes is confined to rural areas \parencite{paxson2010, fernald2011}.} The impact on mental health, which remains unexplored for the HDG in Ecuador, is also a pathway worth considering \parencite{zimmerman2021}.

As noted in Section~\ref{Section 2}, other channels through which the grant could improve future socio-economic outcomes might exist. First, the benefit may alleviate credit constraints, allowing families to invest in productive assets. Although we lack evidence on this mechanism for the Ecuadorian context, the experience of other countries points to its plausibility \parencite{martinez2005, maluccio2010, gertler2012, blattman2020, gelders2019}. The absence of negative effects of the HDG on adult work in the short and medium run \parencite{araujo2017,bosch2019} suggests the feasibility of this channel in relation to household saving and investment behaviour (e.g., facilitating business start-ups or access to assets that enhance employability, such as a vehicle or a driving licence).

A final complementary pathway is an improvement in household resource allocation, particularly when women are the recipients of the transfers. This mechanism is consistent with the findings of \textcite{schady2008b} and \textcite{nabernegg2012}, which indicate that the HDG supports consumption of child-related rather than undesirable goods and services (such as tobacco or alcohol).

\FloatBarrier
\section{Conclusion}\label{Section 5}

Despite having become flagship social programmes in several countries, empirical evidence on the impact of CCTs on long-term outcomes---those more closely aligned with their core objectives than the short-term effects---remains scarce. By targeting poor households with children and promoting human capital formation, policymakers expect these social benefits to improve young adults' income-generating capacity and, in turn, enhance the well-being of future generations.

This paper addresses this gap by providing a rigorous assessment of the effect of Ecuador’s HDG on eligible children’s subsequent formal labour market outcomes. Focusing on individuals aged 26--27, approximately 15 years after initial eligibility, we find that the overall impact of the transfer is sizeable, increasing the likelihood of working in the formal sector by almost 13\%, thereby mitigating the intergenerational transmission of poverty through higher future formal employment and earnings. These findings are consistent with recent evidence for other Latin American and Caribbean countries. As discussed in the paper, for several reasons (ranging from sample selection to selective migration), our estimates are very likely to represent a lower bound of the true effect of the HDG on formal labour market outcomes. Similarly, it is worth noting that we only consider children aged 13--16 who were exposed to the programme for 2--3 years. Focusing on exposure at earlier ages or for longer durations could reveal larger effects than those reported here.

Regarding the external validity of our findings, we should bear in mind that we estimate only the local average intensive margin ITT effects of the HDG. In other words, our results apply only to the neighbourhood of the specific cut-off point in 2008/2009. It is possible that we might observe larger impacts for households with a lower socio-economic status. In fact, previous work has revealed that children living in the country's poorest households are those who experience the largest short-run impacts (in terms of school enrolment and health) of the HDG \parencite{paxson2010,oosterbeek2008,schady2008a}.\footnote{As we are only able to leverage information for observations corresponding to a poverty index between 30.5 and 42.5, we cannot extrapolate ITT effects \enquote{away from the cut-off point}.}

Finally, we also contribute to the literature on the role of conditionality. Unfortunately, we do not have a reliable strategy to identify the causal effect of the weak enforcement of the conditions attached to HDG receipt. Nevertheless, our findings may inform the debate, as prior studies largely examine settings where conditionality is more stringent than in Ecuador \parencite{baird2013}. Similar to these studies, we find a positive impact on long-term outcomes, which suggests that strong conditionality is not necessary for CCTs to fulfil their objectives.

\clearpage
\singlespacing
\printbibliography
\clearpage

\setcounter{table}{0}
\setcounter{figure}{0}
\setcounter{page}{1}
\renewcommand\thetable{A\arabic{table}}
\renewcommand\thefigure{A\arabic{figure}}
\renewcommand\thepage{A\arabic{page}}

\appendix
\section*{Appendix}

\vspace*{\fill}
\begin{figure}[ht]
	\footnotesize
	\centering
	\captionsetup{width=0.75\textwidth}
	\caption{Effects of the discontinuity on the proportion of individuals with a valid identity card number}
	\label{Figure A1}
	\centering
	\noindent\makebox[\textwidth][c]{
		\noindent\begin{minipage}[c]{0.75\textwidth}	
			\centering
			\includegraphics[width=1\textwidth]{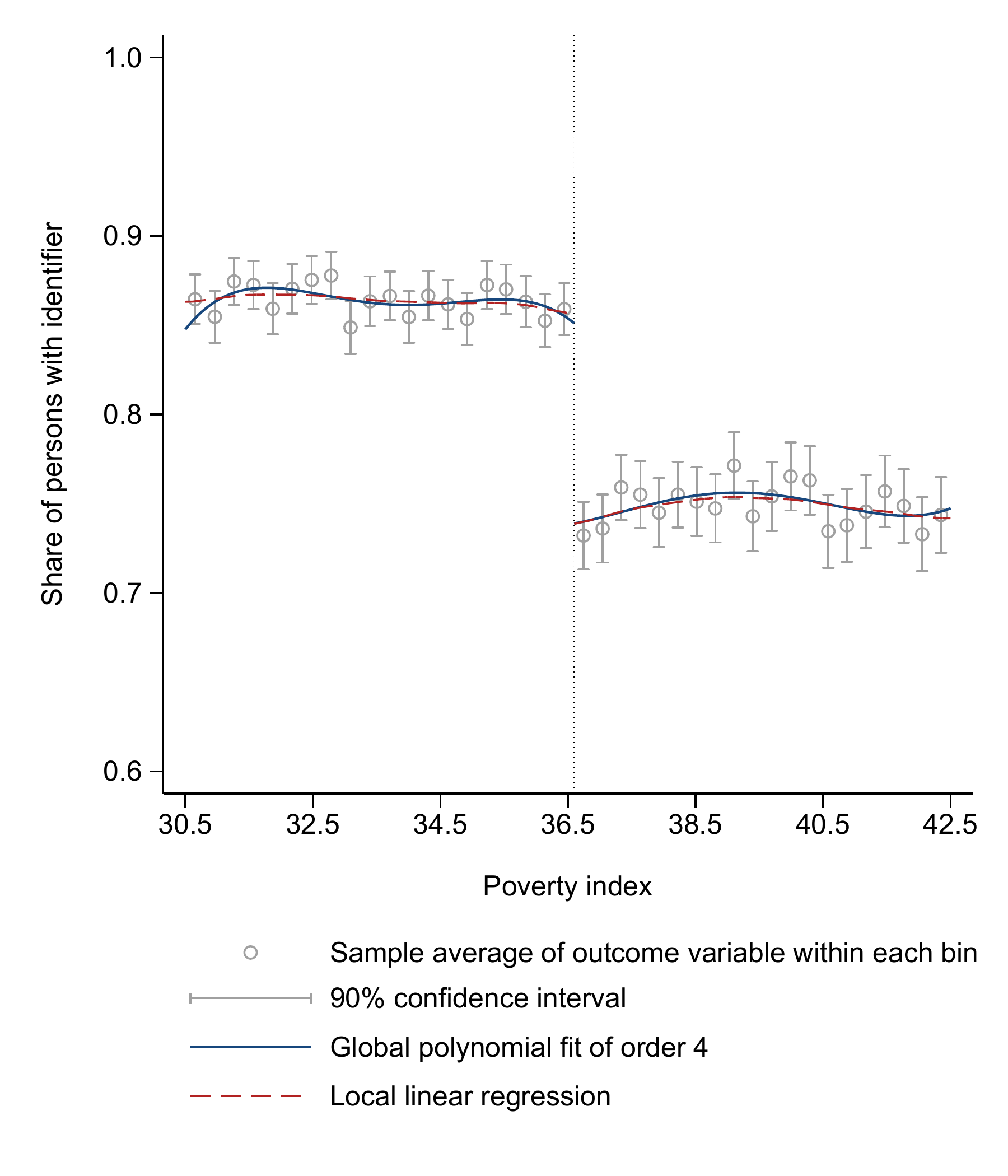} \\
			\justifying
			\noindent\textit{Notes}: The figure shows (i) the sample proportion of individuals with a valid identity card number within each bin and its 90\% confidence interval (grey circles), (ii) the fits of global fourth-order polynomial regressions estimated separately at both sides of the cut-off (blue line) and (iii) the fits of local linear regressions estimated separately on either side of the cut-off using a triangular kernel and data-driven MSE-optimal bandwidths (red line).\\   
			\noindent\textit{Source}: Authors' analysis from \textcite{socialregistry2025} and \textcite{socialsecurity2025}.
		\end{minipage}
	}		
\end{figure}
\vspace*{\fill}
\clearpage

\begin{figure}[p]
	\footnotesize
	\centering
	\captionsetup{width=0.75\textwidth}	
	\caption{Test of continuity in the density of the forcing variable at the cut-off (full sample)}
	\label{Figure A2}
	\centering 
	\noindent\makebox[\textwidth][c]{
		\noindent\begin{minipage}[c]{0.75\textwidth}	
		\centering		
		\includegraphics[width=1\textwidth]{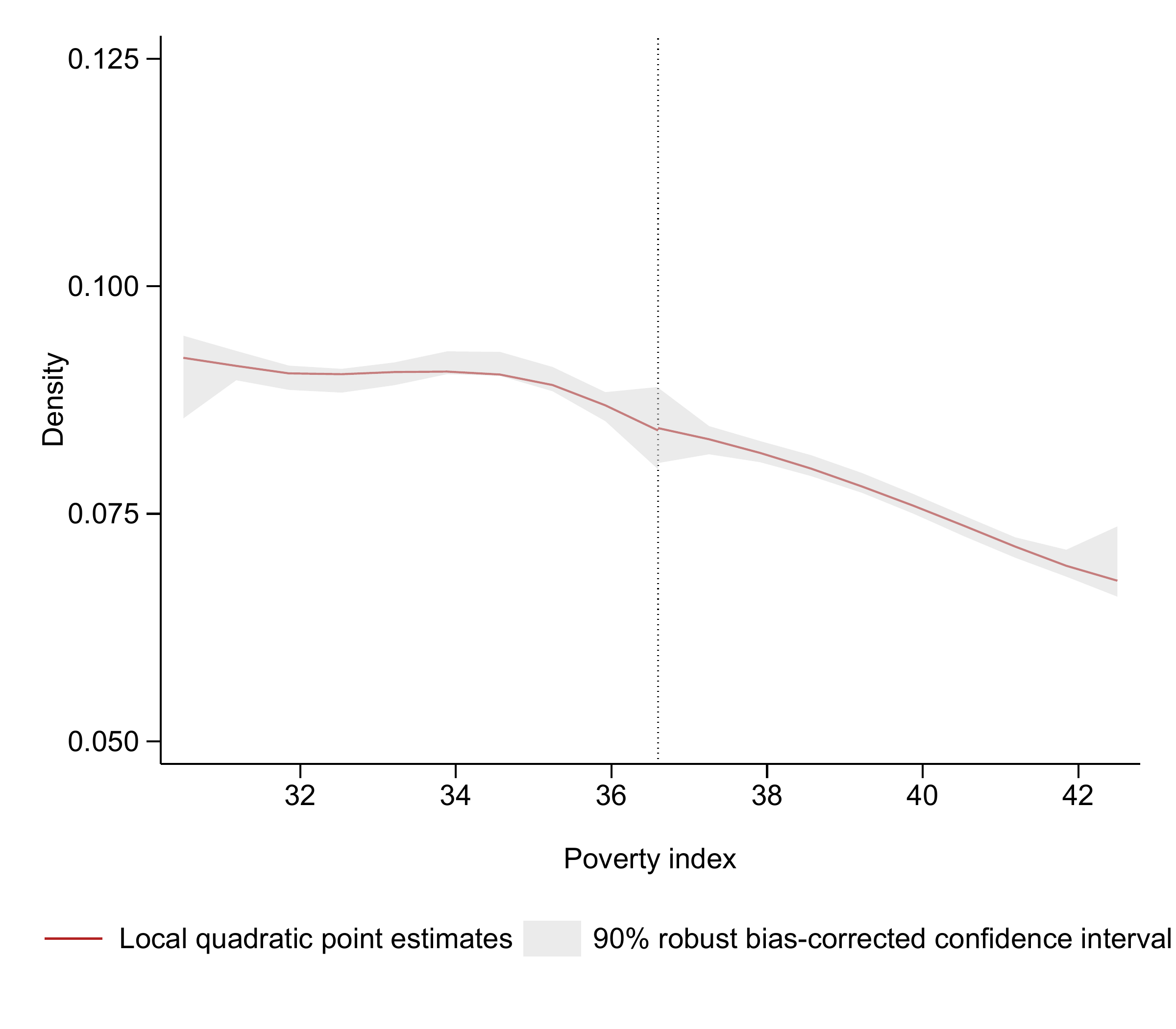} \\
		\justifying
		\noindent\textit{Notes}: The plot explores whether there is a discontinuity in the density of the poverty index using local quadratic regressions estimated separately on either side of the cut-off using a triangular kernel and data-driven MSE-optimal bandwidths. The results of the test do not allow us to reject the hypothesis of continuity ($p-\text{value} = 0.988$).\\   
		\noindent\textit{Source}: Authors' analysis from \textcite{socialregistry2025} and \textcite{socialsecurity2025}.
	\end{minipage}
	}		
\end{figure}	
\clearpage

\begin{figure}[p]
	\footnotesize
	\centering
	\captionsetup{width=0.75\textwidth}	
	\caption{Test of continuity in the density of the forcing variable at the cut-off (complete case-sample)}
	\label{Figure A3}
	\centering 
	\noindent\makebox[\textwidth][c]{
		\noindent\begin{minipage}[c]{0.75\textwidth}	
			\centering		
			\includegraphics[width=1\textwidth]{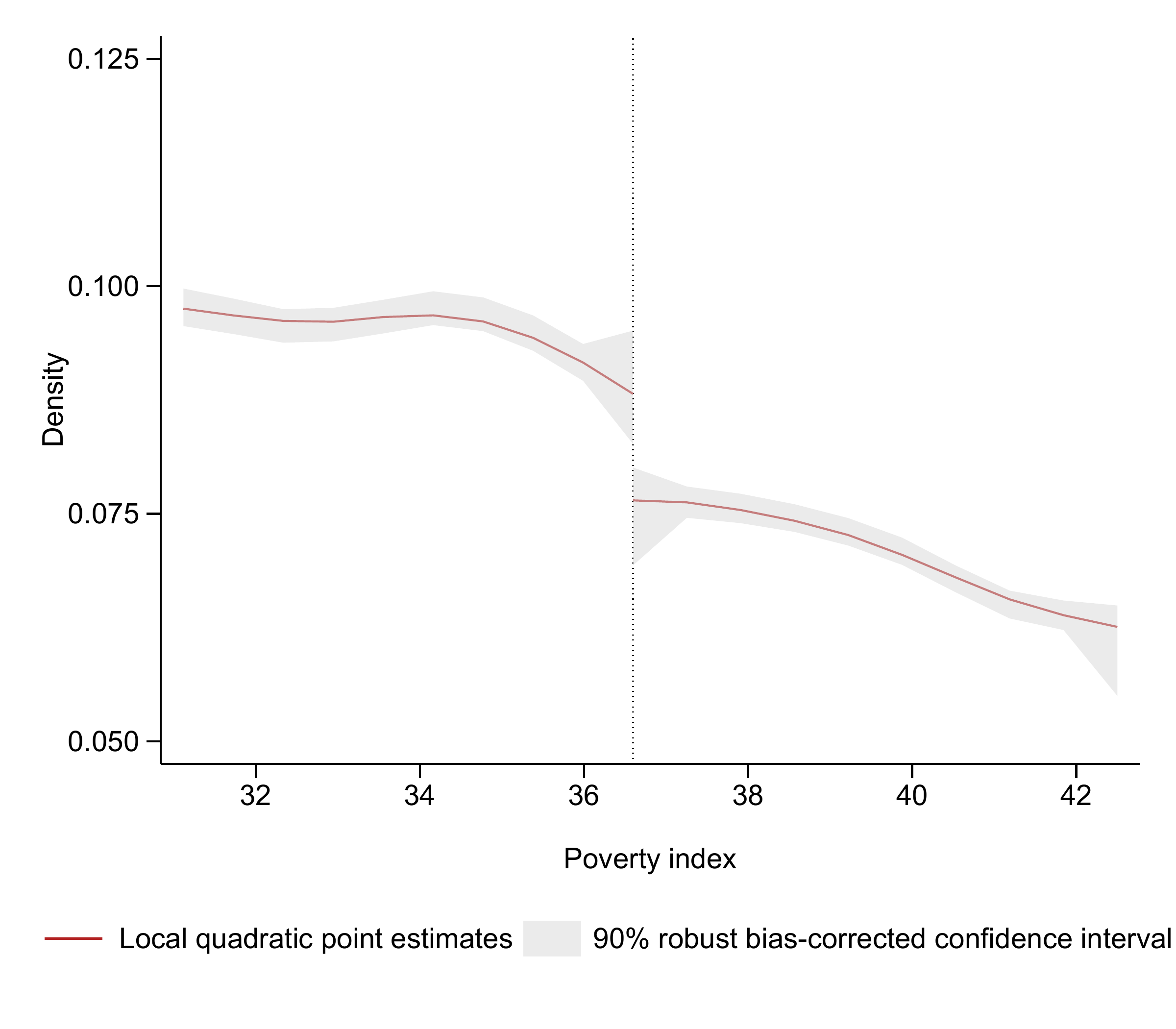} \\
			\justifying
			\noindent\textit{Notes}: The plot explores whether there is a discontinuity in the density of the poverty index using local quadratic regressions estimated separately on either side of the cut-off using a triangular kernel and data-driven MSE-optimal bandwidths. The results of the test allow us to reject the hypothesis of continuity ($p-\text{value} = 0.000$).\\    
			\noindent\textit{Source}: Authors' analysis from \textcite{socialregistry2025} and \textcite{socialsecurity2025}.
		\end{minipage}
	}		
\end{figure}	
\clearpage

\begin{singlespace}
	\begin{table}[p]
		\begin{ThreePartTable}
			\def\sym#1{\ifmmode^{#1}\else\(^{#1}\)\fi}
			\scriptsize
			\setlength\tabcolsep{7pt}
			\sisetup{
					table-number-alignment = center,
					range-phrase = {, }, 
					}
			\begin{TableNotes}[flushleft]\setlength\labelsep{0pt}\footnotesize\justifying
				\item\textit{Notes}: \sym{***} significant at 1\% level; \sym{**} significant at 5\% level; \sym{*} significant at 10\% level. Statistical significance is based on robust bias-corrected inference. Standard errors are clustered at the household level. 90\% confidence intervals based on robust bias-corrected inference are in parentheses. The table shows the results of local linear regressions with the covariate in the heading as the outcome, estimated using a triangular kernel and data-driven MSE-optimal bandwidths. All variables refer to 2008/2009, with the exception of child age (measured in January 2010). The total number of observations in all regressions is \num{59299}.
				\item\textit{Source}: Authors' analysis from \textcite{socialregistry2025} and \textcite{socialsecurity2025}.				
			\end{TableNotes}
			\begin{tabularx}{\linewidth}{X *{3}{S[table-column-width=2cm]}}
					\caption{Covariate balance: evaluation of the discontinuity in the covariates at the cut-off (full sample)} \label{Table A1}\\
					\toprule
					&\multicolumn{1}{c}{RD estimate}&\multicolumn{1}{c}{\makecell{90\% confidence\\interval}}&\multicolumn{1}{c}{\makecell{No. of\\ observations\\effectively used}}\\
					\midrule
					Female&-0.026\sym{*}&[\numrange{-0.057}{-0.001}]&\multicolumn{1}{c}{23244}\\
Ethnic minority&0.004&[\numrange{-0.017}{0.023}]&\multicolumn{1}{c}{23966}\\
Age&-0.015&[\numrange{-0.031}{0.005}]&\multicolumn{1}{c}{15823}\\
Enroled in education&-0.005&[\numrange{-0.018}{0.008}]&\multicolumn{1}{c}{15972}\\
Attending to school&-0.006&[\numrange{-0.019}{0.008}]&\multicolumn{1}{c}{17000}\\
Female household head&-0.061\sym{***}&[\numrange{-0.099}{-0.029}]&\multicolumn{1}{c}{15014}\\
Household head's age&0.034&[\numrange{-0.738}{0.592}]&\multicolumn{1}{c}{13827}\\
Household head's married or in union&0.017&[\numrange{-0.012}{0.057}]&\multicolumn{1}{c}{14188}\\
Household head's years of education&-0.125&[\numrange{-0.367}{0.072}]&\multicolumn{1}{c}{17673}\\
Employed household head&-0.016&[\numrange{-0.040}{0.003}]&\multicolumn{1}{c}{15449}\\
Household head affiliated to Social Security&0.012&[\numrange{-0.010}{0.035}]&\multicolumn{1}{c}{21291}\\
Household size&0.060&[\numrange{-0.035}{0.198}]&\multicolumn{1}{c}{13561}\\
No. of children aged 15 or younger&0.005&[\numrange{-0.060}{0.098}]&\multicolumn{1}{c}{16583}\\
No. of adults aged 60 or older&-0.002&[\numrange{-0.018}{0.014}]&\multicolumn{1}{c}{18415}\\
Household with migrants&0.003&[\numrange{-0.007}{0.017}]&\multicolumn{1}{c}{17538}\\
Rural area&0.020&[\numrange{-0.004}{0.053}]&\multicolumn{1}{c}{16269}\\
Housing is a house or a department&0.042\sym{**}&[\numrange{0.010}{0.075}]&\multicolumn{1}{c}{16989}\\
Concrete or tile roof&0.012&[\numrange{-0.011}{0.041}]&\multicolumn{1}{c}{17529}\\
Tile or parquet floor&0.001&[\numrange{-0.016}{0.019}]&\multicolumn{1}{c}{16663}\\
Concrete, brick or cement walls&0.013&[\numrange{-0.005}{0.035}]&\multicolumn{1}{c}{22744}\\
Roof in good condition&-0.007&[\numrange{-0.026}{0.021}]&\multicolumn{1}{c}{18631}\\
Floor in good condition&0.008&[\numrange{-0.014}{0.036}]&\multicolumn{1}{c}{15877}\\
Walls in good condition&-0.002&[\numrange{-0.021}{0.028}]&\multicolumn{1}{c}{17067}\\
Owner-occupied housing&0.018&[\numrange{-0.008}{0.053}]&\multicolumn{1}{c}{19460}\\
No. of rooms&0.038&[\numrange{-0.030}{0.117}]&\multicolumn{1}{c}{19717}\\
Water supply from public network&-0.013&[\numrange{-0.038}{0.013}]&\multicolumn{1}{c}{21040}\\
Water supply inside the dwelling&-0.002&[\numrange{-0.041}{0.033}]&\multicolumn{1}{c}{13667}\\
Treated drinking water&0.013&[\numrange{-0.012}{0.046}]&\multicolumn{1}{c}{17401}\\
Flush toilet connected to sewer&0.001&[\numrange{-0.027}{0.027}]&\multicolumn{1}{c}{24576}\\
Shower exclusive to the household&-0.004&[\numrange{-0.033}{0.028}]&\multicolumn{1}{c}{20165}\\
Waste disposal through public or private service&-0.004&[\numrange{-0.025}{0.015}]&\multicolumn{1}{c}{22841}\\
Modern cooking energy sources&0.001&[\numrange{-0.005}{0.007}]&\multicolumn{1}{c}{15416}\\
Land ownership&0.020\sym{*}&[\numrange{0.001}{0.040}]&\multicolumn{1}{c}{16307}\\
Ownership of draft animals&0.012&[\numrange{-0.020}{0.043}]&\multicolumn{1}{c}{16061}\\
Remittance-receiving household&0.005&[\numrange{-0.006}{0.019}]&\multicolumn{1}{c}{16445}\\
Household borrowed money&-0.011&[\numrange{-0.032}{0.010}]&\multicolumn{1}{c}{23207}\\
TV ownership&-0.020&[\numrange{-0.044}{0.005}]&\multicolumn{1}{c}{20074}\\
Fridge ownership&-0.014&[\numrange{-0.045}{0.012}]&\multicolumn{1}{c}{22851}\\
DVD ownership&0.018&[\numrange{-0.008}{0.056}]&\multicolumn{1}{c}{12575}\\
Car ownership&0.008&[\numrange{-0.001}{0.020}]&\multicolumn{1}{c}{21374}\\
Washing machine ownership&-0.002&[\numrange{-0.016}{0.009}]&\multicolumn{1}{c}{23033}\\
Computer ownership&0.003&[\numrange{-0.006}{0.010}]&\multicolumn{1}{c}{16200}\\
Cellphone ownership&0.016&[\numrange{-0.016}{0.054}]&\multicolumn{1}{c}{14958}\\
Microwave or oven ownership&0.004&[\numrange{-0.003}{0.012}]&\multicolumn{1}{c}{19063}\\
Blender ownership&0.000&[\numrange{-0.026}{0.028}]&\multicolumn{1}{c}{24520}\\
Cable TV&0.003&[\numrange{-0.005}{0.011}]&\multicolumn{1}{c}{20118}
\\
					\bottomrule	
					\insertTableNotes
			\end{tabularx}
		\end{ThreePartTable}
	\end{table}
\end{singlespace}

\begin{singlespace}
	\begin{table}[p]
		\begin{ThreePartTable}
			\def\sym#1{\ifmmode^{#1}\else\(^{#1}\)\fi}
			\scriptsize
			\setlength\tabcolsep{7pt}
			\sisetup{
				table-number-alignment = center,
				range-phrase = {, }, 
			}
			\begin{TableNotes}[flushleft]\setlength\labelsep{0pt}\footnotesize\justifying
				\item\textit{Notes}: \sym{***} significant at 1\% level; \sym{**} significant at 5\% level; \sym{*} significant at 10\% level. Statistical significance is based on robust bias-corrected inference. Standard errors are clustered at the household level. 90\% confidence intervals based on robust bias-corrected inference are in parentheses. The table shows the results of local linear regressions with the covariate in the heading as the outcome, estimated using a triangular kernel and data-driven MSE-optimal bandwidths. All variables refer to 2008/2009, with the exception of child age (measured in January 2010). The total number of observations in all regressions is \num{48157}.
				\item\textit{Source}: Authors' analysis from \textcite{socialregistry2025} and \textcite{socialsecurity2025}.				
			\end{TableNotes}
			\begin{tabularx}{\linewidth}{X *{3}{S[table-column-width=2cm]}}
				\caption{Covariate balance: evaluation of the discontinuity in the covariates at the cut-off (complete-case sample)} \label{Table A2}\\
				\toprule
				&\multicolumn{1}{c}{RD estimate}&\multicolumn{1}{c}{\makecell{90\% confidence\\interval}}&\multicolumn{1}{c}{\makecell{No. of\\ observations\\effectively used}}\\
				\midrule
				Female&-0.007&[\numrange{-0.047}{0.023}]&\multicolumn{1}{c}{15169}\\
Ethnic minority&0.008&[\numrange{-0.015}{0.032}]&\multicolumn{1}{c}{17510}\\
Age&-0.006&[\numrange{-0.027}{0.017}]&\multicolumn{1}{c}{13258}\\
Enroled in education&-0.004&[\numrange{-0.018}{0.010}]&\multicolumn{1}{c}{13346}\\
Attending to school&-0.006&[\numrange{-0.020}{0.008}]&\multicolumn{1}{c}{15308}\\
Female household head&-0.045\sym{**}&[\numrange{-0.083}{-0.008}]&\multicolumn{1}{c}{13587}\\
Household head's age&0.320&[\numrange{-0.512}{0.890}]&\multicolumn{1}{c}{10821}\\
Household head's married or in union&-0.001&[\numrange{-0.033}{0.041}]&\multicolumn{1}{c}{12314}\\
Household head's years of education&-0.116&[\numrange{-0.367}{0.082}]&\multicolumn{1}{c}{15977}\\
Employed household head&-0.019&[\numrange{-0.041}{0.002}]&\multicolumn{1}{c}{12962}\\
Household head affiliated to Social Security&0.015&[\numrange{-0.007}{0.041}]&\multicolumn{1}{c}{19268}\\
Household size&0.037&[\numrange{-0.064}{0.178}]&\multicolumn{1}{c}{12577}\\
No. of children aged 15 or younger&-0.029&[\numrange{-0.100}{0.055}]&\multicolumn{1}{c}{18233}\\
No. of adults aged 60 or older&0.006&[\numrange{-0.008}{0.020}]&\multicolumn{1}{c}{19982}\\
Household with migrants&0.007&[\numrange{-0.004}{0.020}]&\multicolumn{1}{c}{17356}\\
Rural area&0.015&[\numrange{-0.013}{0.050}]&\multicolumn{1}{c}{14274}\\
Housing is a house or a department&0.044\sym{**}&[\numrange{0.013}{0.078}]&\multicolumn{1}{c}{17321}\\
Concrete or tile roof&0.008&[\numrange{-0.016}{0.040}]&\multicolumn{1}{c}{15661}\\
Tile or parquet floor&0.005&[\numrange{-0.013}{0.022}]&\multicolumn{1}{c}{16330}\\
Concrete, brick or cement walls&0.016&[\numrange{-0.006}{0.043}]&\multicolumn{1}{c}{15750}\\
Roof in good condition&-0.005&[\numrange{-0.029}{0.026}]&\multicolumn{1}{c}{14910}\\
Floor in good condition&0.021&[\numrange{-0.003}{0.052}]&\multicolumn{1}{c}{13416}\\
Walls in good condition&0.008&[\numrange{-0.013}{0.042}]&\multicolumn{1}{c}{13646}\\
Owner-occupied housing&0.034&[\numrange{-0.000}{0.072}]&\multicolumn{1}{c}{14604}\\
No. of rooms&0.073&[\numrange{-0.002}{0.157}]&\multicolumn{1}{c}{16962}\\
Water supply from public network&-0.016&[\numrange{-0.046}{0.009}]&\multicolumn{1}{c}{18041}\\
Water supply inside the dwelling&0.015&[\numrange{-0.028}{0.054}]&\multicolumn{1}{c}{11136}\\
Treated drinking water&0.005&[\numrange{-0.021}{0.036}]&\multicolumn{1}{c}{19268}\\
Flush toilet connected to sewer&-0.004&[\numrange{-0.038}{0.028}]&\multicolumn{1}{c}{17379}\\
Shower exclusive to the household&0.012&[\numrange{-0.017}{0.049}]&\multicolumn{1}{c}{16916}\\
Waste disposal through public or private service&-0.002&[\numrange{-0.028}{0.018}]&\multicolumn{1}{c}{18843}\\
Modern cooking energy sources&0.004&[\numrange{-0.004}{0.010}]&\multicolumn{1}{c}{12939}\\
Land ownership&0.023\sym{*}&[\numrange{0.001}{0.045}]&\multicolumn{1}{c}{13490}\\
Ownership of draft animals&0.005&[\numrange{-0.029}{0.038}]&\multicolumn{1}{c}{14933}\\
Remittance-receiving household&0.005&[\numrange{-0.007}{0.020}]&\multicolumn{1}{c}{14307}\\
Household borrowed money&-0.013&[\numrange{-0.038}{0.017}]&\multicolumn{1}{c}{13884}\\
TV ownership&-0.032\sym{**}&[\numrange{-0.056}{-0.010}]&\multicolumn{1}{c}{20126}\\
Fridge ownership&-0.021&[\numrange{-0.051}{0.012}]&\multicolumn{1}{c}{18610}\\
DVD ownership&0.009&[\numrange{-0.022}{0.049}]&\multicolumn{1}{c}{11113}\\
Car ownership&0.006&[\numrange{-0.007}{0.019}]&\multicolumn{1}{c}{15953}\\
Washing machine ownership&-0.011&[\numrange{-0.031}{0.003}]&\multicolumn{1}{c}{13080}\\
Computer ownership&0.002&[\numrange{-0.008}{0.009}]&\multicolumn{1}{c}{17804}\\
Cellphone ownership&0.005&[\numrange{-0.030}{0.046}]&\multicolumn{1}{c}{12762}\\
Microwave or oven ownership&0.004&[\numrange{-0.002}{0.013}]&\multicolumn{1}{c}{15974}\\
Blender ownership&-0.012&[\numrange{-0.039}{0.020}]&\multicolumn{1}{c}{19373}\\
Cable TV&0.005&[\numrange{-0.004}{0.014}]&\multicolumn{1}{c}{16877}
\\
				\bottomrule	
				\insertTableNotes
			\end{tabularx}
		\end{ThreePartTable}
	\end{table}
\end{singlespace}

\begin{landscape}
	\begin{singlespace}
		\begin{table}[p]
			\begin{ThreePartTable}
				\centering
				\def\sym#1{\ifmmode^{#1}\else\(^{#1}\)\fi}
				\footnotesize
				\setlength\tabcolsep{5pt}
				\sisetup{
					table-number-alignment = center,
					range-phrase = {, },
				}
				\begin{TableNotes}[flushleft]\setlength\labelsep{0pt}\footnotesize\justifying
					\item\textit{Notes}: \sym{***} significant at 1\% level; \sym{**} significant at 5\% level; \sym{*} significant at 10\% level. Standard errors are clustered at the household level. 90\% confidence intervals based on robust bias-corrected inference are in parentheses. Column (I) shows the results of a local linear regression estimated with formal labour income (unconditional on labour market participation) as the outcome, using a triangular kernel and data-driven MSE-optimal bandwidth. The Romano–Wolf stepdown adjusted \textit{p}-values for formal employment and formal labour income are 0.005 in both cases, based on \num{10000} bootstrap replications. Column (II) displays the results of a local linear regression estimated with formal employment as the outcome, using a triangular kernel and data-driven MSE-optimal bandwidth, with multiple imputation of missing outcomes and combination using Rubin's rules. The mean of the outcome and the number of observations effectively used correspond to the first imputation. Column (III) presents the results of a local linear regression with formal employment as the outcome, estimated using a triangular kernel and data-driven MSE-optimal bandwidths, weighted by inverse probability weights. Column (IV) shows the results of a local linear regression estimated with employment as the outcome, using a triangular kernel and data-driven CER-optimal bandwidth. Column (V) displays the results of a local linear regression estimated with formal employment as the outcome, using an Epanechnikov kernel and data-driven MSE-optimal bandwidths. Column (VI) shows the results of a local quadratic regression estimated with formal employment as the outcome, using a uniform kernel and data-driven MSE-optimal bandwidths. The RD estimates in column (I), (IV), (V) and (VI) refer to local average intensive margin ITT effects, while the those in column (II) and column (III) correspond to local average ITT effects under the MAR$|X$ assumption. The control variables include child gender, child age, child ethnicity, household head's gender, household head's age, household head's age squared, household head's marital status, household head's educational attainment, household head's employment status, household size and area of residence.
					\item\textit{Source}: Authors' analysis from \textcite{socialregistry2025} and \textcite{socialsecurity2025}.
				\end{TableNotes}
				\begin{tabularx}{\linewidth}{X *{6}{S[table-column-width=2.5cm]}}
					\caption{Robustness checks (I)} \label{Table A3} \\		
					\toprule
					&\multicolumn{1}{c}{(I)}&\multicolumn{1}{c}{(II)}&\multicolumn{1}{c}{(III)}&\multicolumn{1}{c}{(IV)}&\multicolumn{1}{c}{(V)}&\multicolumn{1}{c}{(VI)}\\ [6pt]
					&\multicolumn{1}{c}{\makecell{Formal\\labour\\income}}&\multicolumn{1}{c}{\makecell{MI}}&\multicolumn{1}{c}{\makecell{IPW}}&\multicolumn{1}{c}{\makecell{CER-\\optimal\\ bandwidth}}&\multicolumn{1}{c}{\makecell{Epanechnikov\\kernel}}&\multicolumn{1}{c}{\makecell{Uniform\\kernel}}\\[12pt]						
					\midrule
					ITT effect&27.985\sym{**}&0.032\sym{**}&0.029\sym{*}&0.057\sym{***}&0.035\sym{**}&0.045\sym{**}\\
&(\numrange{9.899}{54.025})&(\numrange{0.006}{0.068})&(\numrange{0.003}{0.065})&(\numrange{0.022}{0.095})&(\numrange{0.011}{0.071})&(\numrange{0.017}{0.086})\\ [1ex]
No. of observations &\multicolumn{1}{c}{\hspace{1.5mm}\num{48157}}&\multicolumn{1}{c}{\hspace{1.5mm}\num{59299}}&\multicolumn{1}{c}{\hspace{1.5mm}\num{48157}}&\multicolumn{1}{c}{\hspace{1.5mm}\num{48157}}&\multicolumn{1}{c}{\hspace{1.5mm}\num{48157}}&\multicolumn{1}{c}{\hspace{1.5mm}\num{48157}}\\
No. of observations effectively used&\multicolumn{1}{c}{\hspace{1.5mm}\num{12599}}&\multicolumn{1}{c}{\hspace{1.5mm}\num{16714}}&\multicolumn{1}{c}{\hspace{1.5mm}\num{14620}}&\multicolumn{1}{c}{\hspace{3mm}\num{8073}}&\multicolumn{1}{c}{\hspace{1.5mm}\num{13316}}&\multicolumn{1}{c}{\hspace{3mm}\num{8870}}\\
Mean of dependent variable&167.770&0.269&0.270&0.274&0.274&0.274\\
Standard deviation of dependent variable&&&&&&\\ [1ex]
Province fixed effects&\multicolumn{1}{c}{\checkmark}&\multicolumn{1}{c}{\checkmark}&\multicolumn{1}{c}{\checkmark}&\multicolumn{1}{c}{\checkmark}&\multicolumn{1}{c}{\checkmark}&\multicolumn{1}{c}{\checkmark}\\
Control variables&\multicolumn{1}{c}{\checkmark}&\multicolumn{1}{c}{\checkmark}&\multicolumn{1}{c}{\checkmark}&\multicolumn{1}{c}{\checkmark}&\multicolumn{1}{c}{\checkmark}&\multicolumn{1}{c}{\checkmark}
\\
					\bottomrule
					\insertTableNotes
				\end{tabularx}
			\end{ThreePartTable}
		\end{table}
	\end{singlespace}
\end{landscape}

\begin{landscape}
	\begin{singlespace}
		\begin{table}[p]
			\begin{ThreePartTable}
				\centering
				\def\sym#1{\ifmmode^{#1}\else\(^{#1}\)\fi}
				\footnotesize
				\setlength\tabcolsep{5pt}
				\sisetup{
					table-number-alignment = center,
					range-phrase = {, },
				}
				\begin{TableNotes}[flushleft]\setlength\labelsep{0pt}\footnotesize\justifying
					\item\textit{Notes}: \sym{***} significant at 1\% level; \sym{**} significant at 5\% level; \sym{*} significant at 10\% level. Standard errors are clustered at the household level. 90\% confidence intervals based on robust bias-corrected inference are in parentheses. Column (I) shows the results of a local quadratic regression estimated with formal employment as the outcome, using a triangular kernel and data-driven MSE-optimal bandwidths. Column (II) displays the results of a local linear regression estimated with formal employment as the outcome, using a triangular kernel and data-driven MSE-optimal bandwidth excluding observations with a poverty index within a distance of 5\% of the bandwidth of the cut-off. Column (III) presents the results of a local linear regression estimated with formal employment as the outcome, using a triangular kernel and a bandwidth equal to 80\% of the data-driven MSE-optimal bandwidth. Column (IV) shows the results of a local linear regression estimated with formal employment as the outcome, using a triangular kernel and a bandwidth equal to 120\% of the data-driven MSE-optimal bandwidth. Column (V) displays the results of a local linear regression with formal employment as the outcome, estimated using a triangular kernel and data-driven MSE-optimal bandwidth only considering the subsample for which the poverty index is below the threshold and using the median of the poverty index in that segment as the cut-off. Column (VI) presents the results of a local linear regression, estimated using a triangular kernel and data-driven MSE-optimal bandwidth only considering the subsample for which the poverty index is above the threshold and using the median of the poverty index in that segment as the cut-off. In all columns, the RD estimates refer to local average intensive margin ITT effects. The control variables include child gender, child age, child ethnicity, household head's gender, household head's age, household head's age squared, household head's marital status, household head's educational attainment, household head's employment status, household size and area of residence.					
					\item\textit{Source}: Authors' analysis from \textcite{socialregistry2025} and \textcite{socialsecurity2025}.
				\end{TableNotes}
				\begin{tabularx}{\linewidth}{X *{6}{S[table-column-width=2.5cm]}}
					\caption{Robustness checks (II)} \label{Table A4} \\		
					\toprule
					&\multicolumn{1}{c}{(I)}&\multicolumn{1}{c}{(II)}&\multicolumn{1}{c}{(III)}&\multicolumn{1}{c}{(IV)}&\multicolumn{1}{c}{(V)}&\multicolumn{1}{c}{(VI)}\\ [6pt]
					&\multicolumn{1}{c}{\makecell{Local\\quadratic\\regression}}&\multicolumn{1}{c}{\makecell{\enquote{Donut-hole}\\approach}}&\multicolumn{1}{c}{\makecell{Narrower\\bandwidth}}&\multicolumn{1}{c}{\makecell{Wider\\bandwidth}}&\multicolumn{1}{c}{\makecell{Placebo\\on the left}}&\multicolumn{1}{c}{\makecell{Placebo\\on the right}}\\[12pt]						
					\midrule
					ITT effect&0.046\sym{**}&0.032\sym{*}&0.051\sym{**}&0.042\sym{**}&0.008&0.016\\
&(\numrange{0.015}{0.084})&(\numrange{0.000}{0.076})&(\numrange{0.023}{0.113})&(\numrange{0.018}{0.092})&(\numrange{-0.030}{0.050})&(\numrange{-0.032}{0.059})\\ [1ex]
No. of observations&\multicolumn{1}{c}{\hspace{1.5mm}\num{48157}}&\multicolumn{1}{c}{\hspace{1.5mm}\num{47482}}&\multicolumn{1}{c}{\hspace{1.5mm}\num{48157}}&\multicolumn{1}{c}{\hspace{1.5mm}\num{48157}}&\multicolumn{1}{c}{\hspace{1.5mm}\num{28069}}&\multicolumn{1}{c}{\hspace{1.5mm}\num{20088}}\\
No. of observations effectively used&\multicolumn{1}{c}{\hspace{1.5mm}\num{23074}}&\multicolumn{1}{c}{\hspace{1.5mm}\num{11727}}&\multicolumn{1}{c}{\hspace{1.5mm}\num{18411}}&\multicolumn{1}{c}{\hspace{1.5mm}\num{27609}}&\multicolumn{1}{c}{\hspace{3mm}\num{8424}}&\multicolumn{1}{c}{\hspace{3mm}\num{7093}}\\
Mean of dependent variable&0.274&0.274&0.274&0.274&0.274&0.274\\ [1ex]
Province fixed effects&\multicolumn{1}{c}{\checkmark}&\multicolumn{1}{c}{\checkmark}&\multicolumn{1}{c}{\checkmark}&\multicolumn{1}{c}{\checkmark}&\multicolumn{1}{c}{\checkmark}&\multicolumn{1}{c}{\checkmark}\\
Control variables&\multicolumn{1}{c}{\checkmark}&\multicolumn{1}{c}{\checkmark}&\multicolumn{1}{c}{\checkmark}&\multicolumn{1}{c}{\checkmark}&\multicolumn{1}{c}{\checkmark}&\multicolumn{1}{c}{\checkmark}
\\
					\bottomrule
					\insertTableNotes
				\end{tabularx}
			\end{ThreePartTable}
		\end{table}
	\end{singlespace}
\end{landscape}

\begin{singlespace}
	\begin{table}[p]
		\begin{ThreePartTable}
			\def\sym#1{\ifmmode^{#1}\else\(^{#1}\)\fi}
			\footnotesize
			\setlength\tabcolsep{5pt}
			\sisetup{
				table-number-alignment = center,
				range-phrase = {, }, 
			}
			\begin{TableNotes}[flushleft]\setlength\labelsep{0pt}\footnotesize\justifying
				\item\textit{Notes}: \sym{***} significant at 1\% level; \sym{**} significant at 5\% level; \sym{*} significant at 10\% level. Statistical significance is based on robust bias-corrected inference. Standard errors are clustered at the household level. 90\% confidence intervals based on robust bias-corrected inference are in parentheses. The table shows the results of local linear regressions with completion of primary education as the outcome, estimated using a triangular kernel and data-driven MSE-optimal bandwidths. In the most comprehensive specification, the Romano–Wolf stepdown-adjusted \textit{p}-values for formal employment, formal labour income and primary education completion are 0.006 for all three outcomes, based on \num{10000} bootstrap replications. The control variables include child gender, child age, child ethnicity, household head's gender, household head's age, household head's age squared, household head's marital status, household head's educational attainment, household head's employment status, household size and area of residence.
				\item\textit{Source}: Authors' analysis from \textcite{socialregistry2025}.				
			\end{TableNotes}
			\begin{tabularx}{0.9\textwidth}{X *{3}{S[table-column-width=2.5cm]}}
				\caption{Local average intensive margin ITT effect of the HDG on the probability of primary education completion in 2013/2014} \label{Table A5}\\
				\toprule
				&\multicolumn{1}{c}{(I)}&\multicolumn{1}{c}{(II)}&\multicolumn{1}{c}{(III)}\\
				\midrule
				ITT effect&0.017\sym{*}&0.016\sym{*}&0.017\sym{*}\\
&[\numrange{0.000}{0.036}]&[\numrange{0.001}{0.035}]&[\numrange{0.002}{0.036}]\\ [1ex]
No. of observations &\multicolumn{1}{c}{\hspace{1.5mm}\num{26255}}&\multicolumn{1}{c}{\hspace{1.5mm}\num{26255}}&\multicolumn{1}{c}{\hspace{1.5mm}\num{26255}}\\
No. of observations effectively used  &\multicolumn{1}{c}{\hspace{3mm}\num{9442}}&\multicolumn{1}{c}{\hspace{1.5mm}\num{10255}}&\multicolumn{1}{c}{\hspace{3mm}\num{9836}}\\
Mean of dependent variable&0.969&0.969&0.969\\ [1ex]
Province fixed effects&&\multicolumn{1}{c}{\checkmark}&\multicolumn{1}{c}{\checkmark}\\
Control variables&&&\multicolumn{1}{c}{\checkmark}
\\
				\bottomrule	
				\insertTableNotes
			\end{tabularx}
		\end{ThreePartTable}
	\end{table}
\end{singlespace}

\end{document}